\RequirePackage{snapshot} 
\documentclass[12pt]{amsart}
\usepackage[all]{xy}
\usepackage[dvipsnames, table]{xcolor}

\usepackage{graphics, hyperref, color,  tikz, pgf, paralist,
    enumitem, ifthen, comment, amsthm, amsmath, amssymb, subcaption, pgfplots, 
    verbatim, float, pgfplotstable, layouts, verbatim, xkeyval, calc,
    booktabs, cancel}
    
\usepackage[tableposition=above]{caption}

\usepackage[colorinlistoftodos]{todonotes}
\usepackage[export]{adjustbox}
\usepgfplotslibrary{groupplots, colormaps}

\usetikzlibrary{shapes, arrows.meta, calendar, matrix, backgrounds, folding,
calc, positioning, spy, decorations.pathreplacing, external, plotmarks}

\graphicspath{{figures/images/jpgs/}}

\usepackage[margin=0.4in]{geometry}

\pgfplotsset{compat=1.9, base/.style={xtick pos=bottom,
																ytick pos=left,
																xlabel style={font=\scriptsize}, 
																ylabel style={font=\scriptsize},
																title style={align=center, font=\footnotesize},
																enlargelimits=false,
																ticklabel style={font=\tiny},
																grid=both,
																grid style={line width=.1pt, draw=gray!10},
																major grid style={line width=.15pt,draw=gray!30},
																legend style={legend pos=north east, font=\tiny}
																},
											convergence/.style = {base,
																				width=.7\textwidth,
																				height=.5\textwidth,
																				enlargelimits=true},
											convergence_narrow/.style = {base,
																				width=.45\textwidth,
																				height=.5\textwidth,
																				enlargelimits=true},
											embedding/.style = {base,
																			 width=.95\textwidth,
																			 height=.5\textwidth},
											wavelength/.style = {base,
																			 width=.95\textwidth,
																			 height=.5\textwidth,
																			 enlargelimits=true,
																		     xlabel={Wavelength ($\mu$m)},
																		     ylabel={Loss (dB/m)},
																			 x tick label style={font=\tiny,
																							  			    /pgf/number format/zerofill,
																							  			    /pgf/number format/precision=2},																		     }
		    								}

\pgfplotsset{legend image code/.code={\draw[mark repeat=2,mark phase=2]
plot coordinates {(0cm,0cm)
							 (0.15cm,0cm)        
							( 0.3cm,0cm)         
							};
							}
							}

\definecolor{BlueViolet}{rgb}{0.541, 0.169, 0.886}

\newcommand{\micrometer}{\, \mu \mbox{m}}

\newcolumntype{g}{>{\columncolor{gray!10}}c}

\definecolor{dgreen}{rgb}{0.2, 0.5, 0.2}

\newcommand{\C}{\mathbb{C}}
\newcommand{\R}{\mathbb{R}}

\newcommand{\ii}{\hat{\imath}}

\newcommand{\St}{\tilde{S}}

\newcommand{\Vt}{\tilde{V}}

\newcommand{\vL}{\varLambda}
\newcommand{\veps}{\varepsilon}

\renewcommand{\Gamma}{\varGamma}
\renewcommand{\Lambda}{\varLambda}
\renewcommand{\vL}{\varLambda}
\newcommand{\vG}{\varGamma}

\newcommand{\og}{\omega}
\newcommand{\om}{\varOmega}
\newcommand{\oh}{\varOmega_h}

\newcommand{\Ho}{\mathring{H}}
\newcommand{\Nedelec}{N{\'{e}}d{\'{e}}lec } 

\renewcommand{\AA}{\mathcal{A}}
\newcommand{\BB}{\mathcal{B}}

\newcommand{\Eth}{E^{h}_\tau}
\newcommand{\vph}{\varphi^{h}}
\newcommand{\vphi}{\varphi}
\newcommand{\Vhp}{\mathcal{V}^{h}}
\newcommand{\Nhp}{\mathcal{N}^{h}}
\newcommand{\nN}{n_\mathcal{N}}
\newcommand{\nV}{n_\mathcal{V}}

\DeclareMathOperator{\rot}{rot}
\DeclareMathOperator{\dive}{div}
\DeclareMathOperator{\curl}{curl}
\DeclareMathOperator{\grad}{grad}
\renewcommand{\d}{\partial}

\def\d{\partial}

\newcommand*{\SetElement}[4]{ \pgfplotstablegetelem{#2}{#3}\of{#1} \let#4\pgfplotsretval}

\theoremstyle{remark}

\author{P.~Vandenberge}
\address{Portland State University, PO Box 751, Portland OR 97207,USA }
\email{piet2@pdx.edu}

\author{J.~Gopalakrishnan}
\address{Portland State University, PO Box 751, Portland OR 97207,USA }
\email{gjay@pdx.edu}

\author{J. Grosek}
\address{Air Force Research Laboratory, 3550 Aberdeen Ave SE, Kirtland Air Force Base,
  NM 87117, USA}
\email{AFRL.RDL.OrgBox@us.af.mil}

\title{Sensitivity of Confinement Losses in Optical Fibers\\ to Modeling Approach}

\begin{document}

\begin{abstract}
%
    
	A prime objective of modeling optical fibers is capturing mode confinement losses correctly. 
  This paper demonstrates that specific modeling choices, especially regarding the outer fiber cladding regions and the placement of the computational boundary, have significant impacts on the calculated mode losses.
  Our results illustrate that one can obtain disparate mode confinement loss profiles for the same optical fiber design simply by moving the boundary to a new material region.
  We conclude with new recommendations for how to better model these losses.
\end{abstract}

\maketitle

\section{Introduction}
\label{sect:intro}

The primary concern of this effort is exploring the effect of modeling choices on the computed 
mode confinement losses (CLs) of microstructured optical fibers.
Many microstuctured fibers guide light by interference and/or anti-resonant reflection effects~\cite{litchinitserAntiresonantReflectingPhotonic2002}.  Such guidance mechanisms are
often inherently lossy, and this loss is modeled by finding so-called leaky modes.   These modes
satisfy an eigenvalue problem derived from Maxwell equations on the fiber's transverse geometry.
The confinement loss can then be found from the imaginary component of the associated eigenvalue. The loss value as a function of input wavelength gives the fiber's
loss profile.
This work shows that loss profiles are sensitive to various modeling choices, 
especially those 
regarding the material in which the boundary of the computational domain is placed.

It seems to be common practice to place this boundary in the innermost cladding region of 
the fiber, effectively allowing the cladding material to extend to infinity~\cite[p.~62ff]{marcuseTheoryDielectricOptical2013}.
The assumption underlying this choice is that any material changes or structures beyond this cladding region have little to no effect on the core modes of the given fiber.
This reasoning may be well-justified for optical fibers that guide light by total internal reflection, such as  step-index or graded-index fiber designs, where the guided mode profiles exponentially decay in the radial direction outside of the core region.
In contrast, leaky modes do not necessarily decay exponentially in the cladding \cite{belardiHollowAntiresonantFibers2014}.  Furthermore, all leaky modes
eventually grow exponentially as we move radially
away from the fiber core. Therefore, for leaky modes, unlike
guided modes, exponential decay cannot be used to justify a model with
infinite cladding.  

Premature termination of the computational domain can have
significant effects on the loss profile of a given fiber.
Bragg Fibers, which consist of concentric rings of dielectric material, referred to as ``anti-resonant layers'' in~\cite{birdAttenuationModelHollowcore2017}, have resonant wavelengths at 
which their losses spike \cite{duguayAntiresonantReflectingOptical1986a}.  The location
of these
wavelengths depend on the configuration of these concentric rings, and may occur 
on several scales, dependent on the thickness of the layers.  During modeling, if 
the computational boundary truncates what would otherwise be a full anti-resonant layer,
an entire set of loss peaks may be removed from the spectral profile.

\begin{figure}
  \centering
 \newcommand{\Ntubes}{7}

\newcommand{\scale}{.75}
\newcommand{\Rcore}{\scale*30pt}
\newcommand{\Tclad}{\scale*24pt}
\newcommand{\Tpoly}{\scale*40pt}

\newcommand{\Rcap}{\scale*16pt}
\newcommand{\Tcap}{\scale/.85*2pt}

\newcommand{\magnify}{13}

\newcommand{\embfrac}{.4}
\newcommand{\Rclad}{\Rcore+\Tcap+\Rcap + \Tcap - \embfrac*\Tcap+\Rcap}
\newcommand{\Rtube}{\Rcore+\Tcap+\Rcap}
\newcommand{\Rtotal}{\Rclad+\Tclad+\Tpoly}

\pgfdeclarelayer{fg}    
\pgfsetlayers{main,fg}  

	\begin{tikzpicture} [spy using outlines, ultra thin]

       \fill [blue!8,even odd rule,opacity=1] (0,0)
       circle[radius=\Rclad + \Tclad + \Tpoly]
       circle[radius=\Rclad + \Tclad];

      \fill [blue!15,even odd rule,opacity=1] (0,0)
      circle[radius=\Rclad + \Tclad]
      circle[radius=\Rclad];

	  \draw[BrickRed, dashed, line width = .4pt] (0,0)
	  circle[radius=.98*\Rcore];
	  
      \draw[BrickRed!50!red, - stealth, line width = .4pt] (0,0) -- (0:\Rcore)
      node [ black, pos=.35, below]
      {\scriptsize $R_{\mbox{\tiny core}}$};
      
      \draw[BrickRed!50!red, | - |, line width = .4pt] (\Rclad + .02*\Tclad, 0) - - (\Rclad + .98*\Tclad, 0)
      node [black, midway, below]
      {\scriptsize $t_{\mbox{\tiny clad}}$};
      
      \draw[BrickRed!50!red, | - |, line width = .4pt] (\Rclad+1.01*\Tclad, 0)--++(\Tpoly, 0)
      node [black, midway, below]
      {\scriptsize $t_{\mbox{\tiny poly}}$};
      
    \foreach \a in {1,...,\Ntubes}
    {
    	\fill [blue!35,even odd rule,opacity=.7] (\a*360/\Ntubes + 90: \Rcore+\Tcap+\Rcap) 
      circle [radius=\Rcap+\Tcap]
      circle [radius=\Rcap];
    }

      \spy [black, draw, height=5*\Rcore,width=7*\Rcore, magnification=\magnify, connect spies] on (0, \Rclad) in node[ultra thin] at (6*\Rtotal, 0);
      
    \begin{pgfonlayer}{fg}
    
    \draw[- stealth, line width=.8pt, BrickRed!50!red] (6*\Rtotal ,-0.8cm) -- (6*\Rtotal,0) ;
    \draw[- stealth, line width=.8pt, BrickRed!50!red] (6*\Rtotal, \embfrac* \Tclad + .8cm) -- (6*\Rtotal, 1.1*\embfrac* \Tclad) node [black, pos=0.12,above] {\scriptsize $e_{\mbox{\tiny cap}}$};
        
    \coordinate (A) at ($(6*\Rtotal, 0) - (0, \magnify*\Rclad) + (0, \magnify*\Rtube) - (0, \magnify*\Rcap)+(75: \magnify*\Rcap)$);
    \coordinate (B) at ($(6*\Rtotal,0)+(-90:\magnify*\Rclad)+(90:\magnify*\Rtube)+(-90: \magnify*\Rcap)+(75:\magnify*\Rcap+ .95*\magnify*\Tcap)$) ;

    \draw[BrickRed!50!red, | - |, line width = .6pt] (A) -- (B) node [black, pos=0, below] {\scriptsize $t_{\mbox{\tiny cap}}$};;

      \end{pgfonlayer}
             
    \end{tikzpicture}
  \caption{Geometry of a typical anti-resonant fiber (ARF).}
\label{fig: arf geom}
\end{figure}
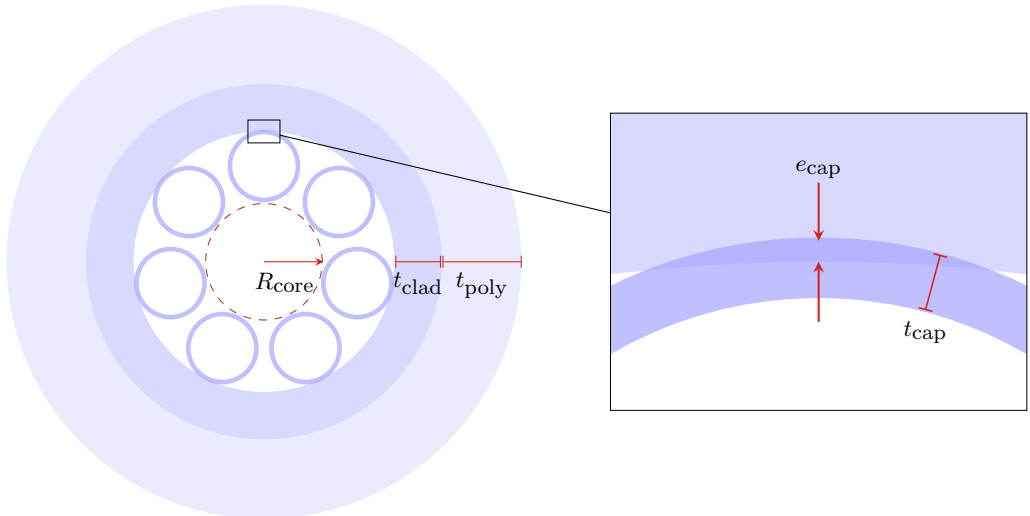

In order to elucidate these possible effects on modern optical waveguides, 
we model two anti-resonant fibers (ARFs) from
the literature \cite{polettiNestedAntiresonantNodeless2014,
kolyadinLightTransmissionNegative2013} and compare their spectral loss curves
for different choices of boundary placement.  We will see that placing the
computational boundary beyond the cladding (in a lower index ambient material)
increases the variability in the loss profiles dramatically, to levels beyond
that seen experimentally.  Contrasting this, the loss profiles
of these same fibers are smooth when the computational domain is placed in the
cladding. We then show that including a lossy  material layer, corresponding to
polymer in the real life fiber, 
smooths out some this extreme variability, 
suggesting that knowledge of the material loss properties of fiber coating
 materials may be important for fine grained
modeling of these devices. 

Such fine grained modeling is of particular importance for microstructured
fibers. The microstructure of the anti-resonant fibers considered here consists
of capillary tubes attached at regular angular intervals to an outer cladding.  Varying
the small distance of separation between these capillary tubes
affects the loss values by up to three orders of magnitude (see 
\cite[Figure~6]{polettiNestedAntiresonantNodeless2014}).
In this work, we detail the
results of a numerical study on what we term the ``embedding distance'' of the
capillary tubes, which quantifies what portion of the tube wall has melded with
the cladding (see Figure~\ref{fig: arf geom} inset).  This study involves similarly small distances, with variation in tube
placement at most $0.42 \micrometer$. The resulting losses
depends strongly on the model's outer material configuration, with variation up
to three orders of magnitude in one configuration and no variation in another. 
This suggests that some loss mechanisms present in the fiber may be
missed by models placing the computational boundary exclusively in the cladding.

{\em Outline.} The rest of this paper is organized as follows.
In Section~\ref{sec:bragg-fiber}, we 
leverage semi-analytic solutions for Bragg fibers
 to establish the effects of outer material configuration on the core modes, 
 and to verify the accuracy of our numerical mode solver.
Then, in Section~\ref{sec:arf-CL}, two anti-resonant fibers from the
 literature \cite{polettiNestedAntiresonantNodeless2014, 
 kolyadinLightTransmissionNegative2013} are examined, showing that their 
 spectral mode loss curves are highly sensitive to different choices of 
 boundary placement. Next, in
Section~\ref{sec:polymer}, we repeat these studies incorporating an
additional lossy polymer layer into the model.  In Section~\ref{sec:emb},
we investigate the sensitivity of CL to above-mentioned embedding
distance for all choices of outer material configurations.
Next, in Section~\ref{sec:accel-conv}, we exploit the understanding
gained in the previous sections to show that informed meshing practices
can accelerate convergence of numerical results.
We discuss the implications of these results for
modeling microstructured fibers in Section~\ref{sec:discussion}.
Finally, our Appendix~\ref{sec:numerical-method} describes the eigenproblem
 that is solved to produce the mode profiles and their corresponding 
 propagation constants and loss values. 
We there also describe the FEAST algorithm as it applies to solving
generalized eigenvalue problems, 
as it provides multiple propitious qualities for mode finding and 
achieving mathematically converged results.

\section{Bragg Fiber and Numerical Verification}
\label{sec:bragg-fiber}

In this section, we show the effects of changing the outer material configuration on 
Bragg fibers, and verify that we can capture these changes using our numerical methods.
We also note that Bragg fibers are used as reference fibers for ARFs, and give 
the loss profiles for the reference fibers for the ARFs we are studying.


\begin{figure}
     \centering
     \begin{subfigure}[b]{0.35\textwidth}
         \centering
         \includegraphics[width=.85\textwidth]{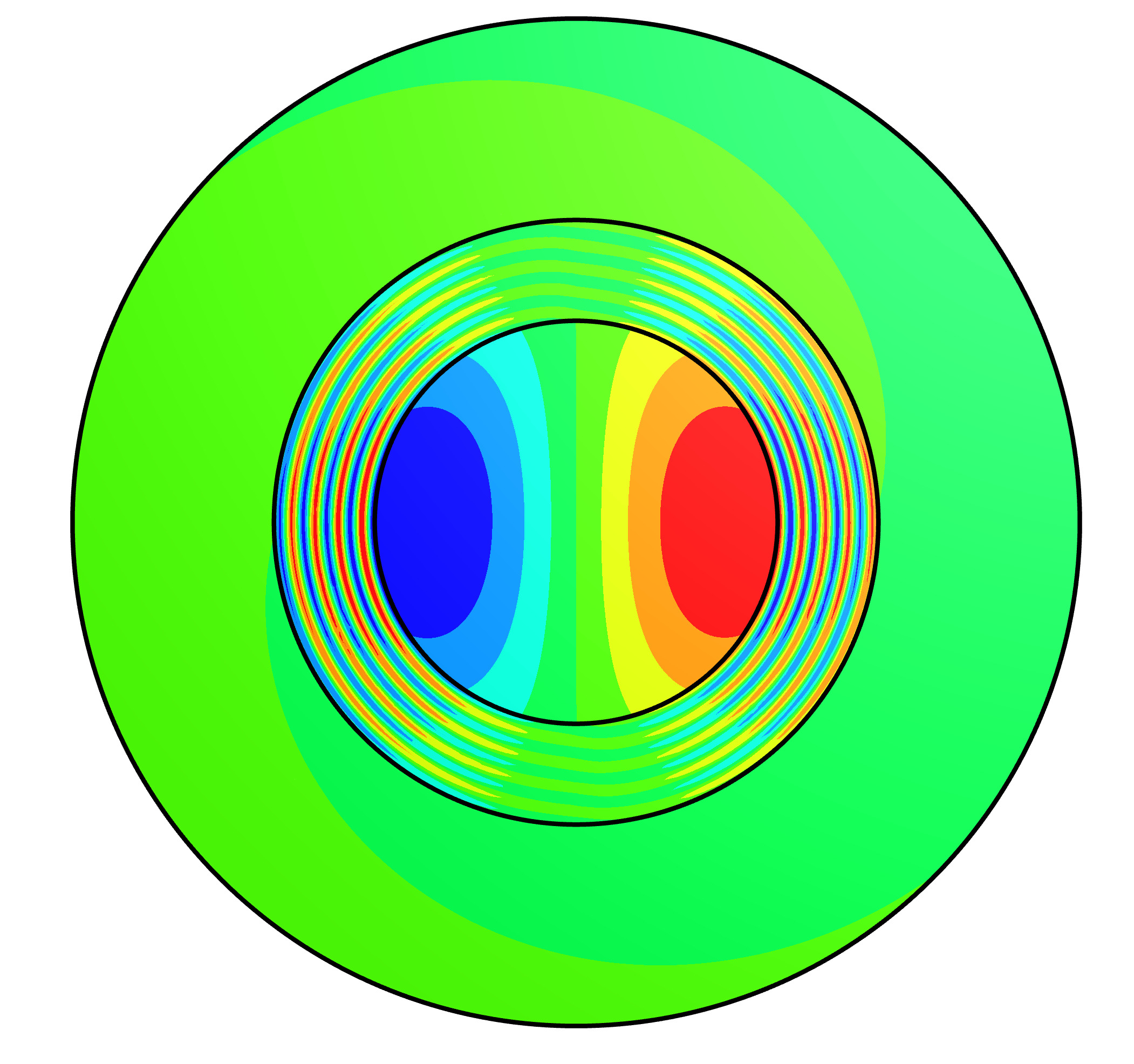}
         \caption{Longitudinal component}
         \label{fig:bragg-longitudinal}
     \end{subfigure}
     \hspace{.2in}
     \begin{subfigure}[b]{0.35\textwidth}
         \centering
         \includegraphics[width=.85\textwidth]{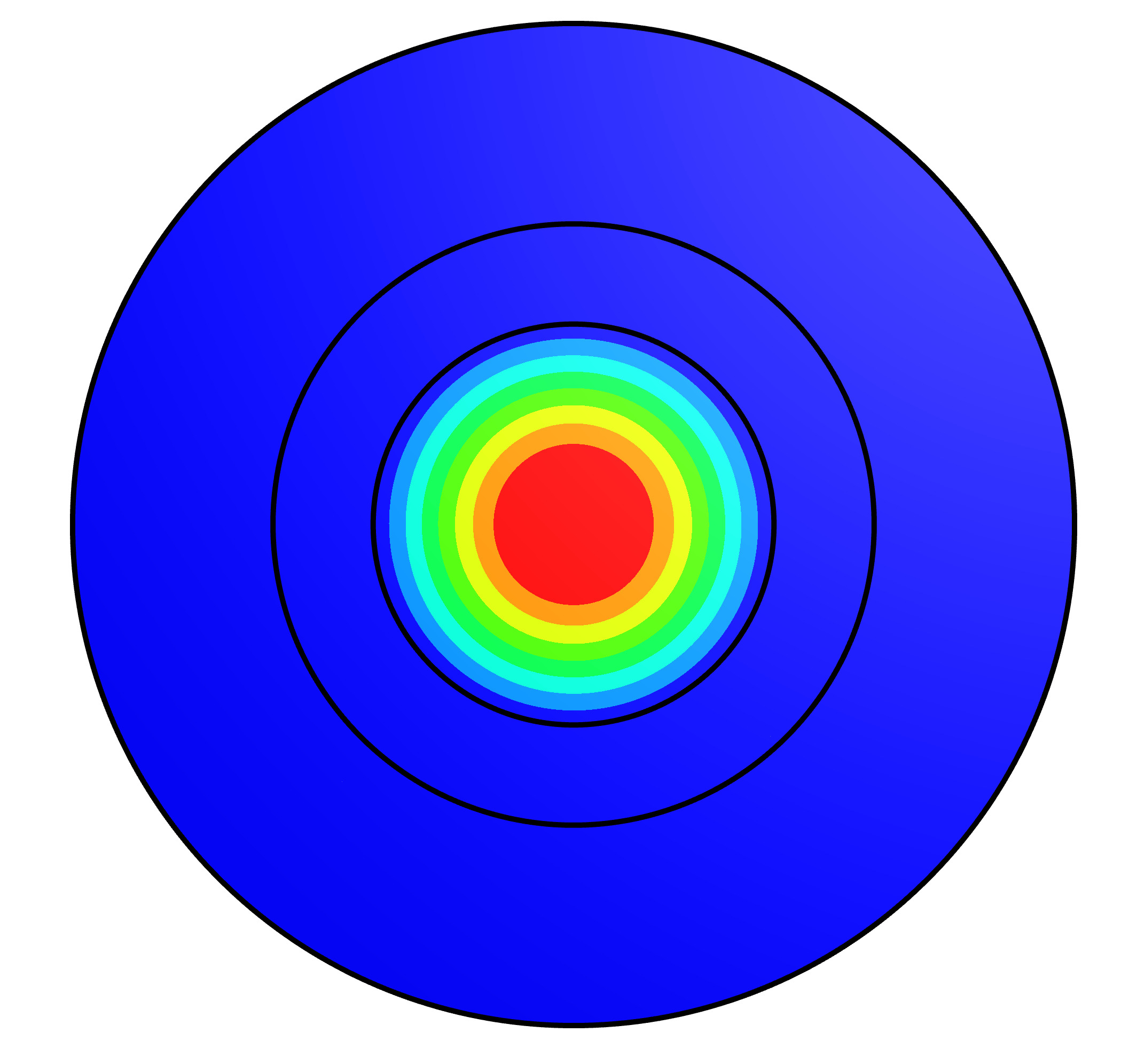}
         \caption{Transverse field magnitude}
         \label{fig:bragg-transverse}
     \end{subfigure}
     \par\medskip
     \begin{subfigure}[b]{0.35\textwidth}
         \centering
         \includegraphics[width=1\textwidth]{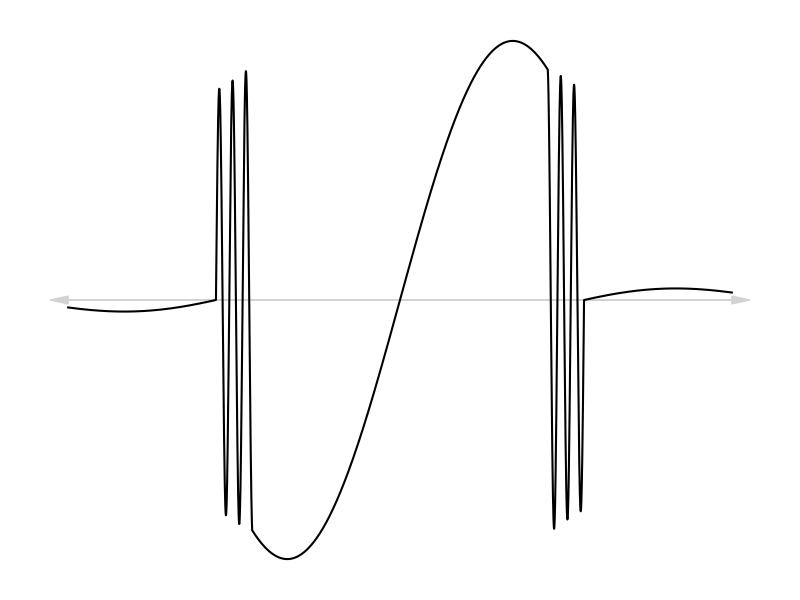}
         \caption{Cross section of Fig.~\ref{fig:bragg-longitudinal} profile}
     \end{subfigure}
     \hspace{.45in}
     \begin{subfigure}[b]{0.35\textwidth}
         \centering
         \includegraphics[width=1\textwidth]{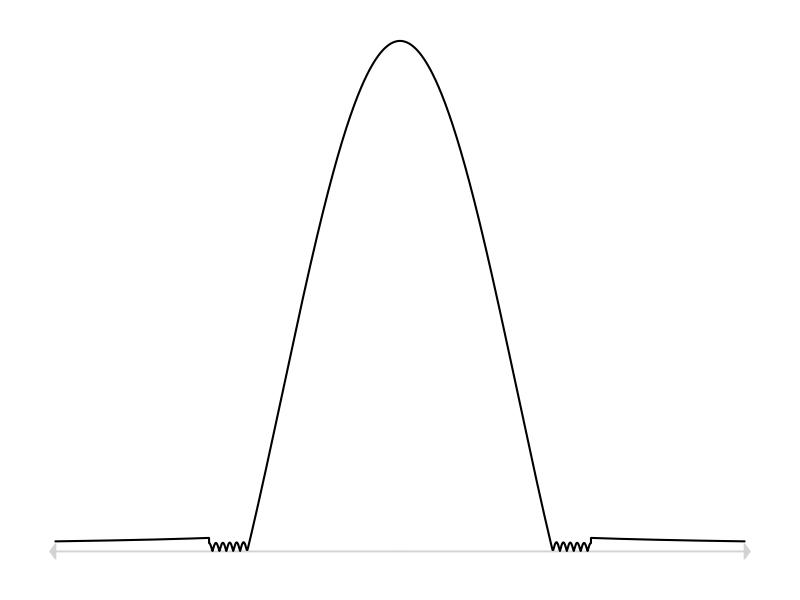}
         \caption{Cross section of Fig.~\ref{fig:bragg-transverse} profile}
     \end{subfigure}
     \caption{Bragg $N_1$ Fiber Fundamental Mode Profiles and their 1-D centered horizontal cross sections}
\label{fig: bragg mode profiles}
\end{figure}

It was shown in \cite{yehBraggReflectionWaveguides1976} that a waveguide with a
low index core region could support lossless modes provided that the cladding
consists of periodic high and low index material. The Bragg Fiber is an optical
fiber design that implements this insight.  It is radially symmetric (azimuthaly invariant) and formed of concentric rings (annuli) of homogeneous material, with the final ambient material around the fiber extending to infinity.  The necessity of terminating the infinitely periodic cladding of the theoretical fiber introduces some loss.  
The analytic calculations used here for finding this loss are based on the matrix based methodology established in the 1978 publication~\cite{yehTheoryBraggFiber1978}.

%

Throughout this work, we utilize the terminology in \cite{birdAttenuationModelHollowcore2017}: $N_i$ denotes a Bragg fiber
with $i$ number of complete (or finite-width) anti-resonant layers. 
Thus, a hollow core surrounded by a high index
material represents the $N_0$ fiber, and the $N_1$ fiber design is a glass ring suspending in low index material, et cetera.
The low index material is air with refractive index $n_{air} = 1.00027717$ and the high
index material is silica glass with refractive index $n_{glass} =
1.4388164768221814$.  The core radius is $r = 40.775 \micrometer$ and
thickness of the glass ring is $t = 10 \micrometer$.  For the $N_2$ fiber, the
thickness of the complete air layer beyond the glass ring was set equal to
the core radius.
Our numerical studies, not shown here, demonstrated that chromatic dispersion does not affect any of our conclusions to this effort, and since including chromatic dispersion necessitates the later use of an extra numerical solve in equation \ref{eq: spikes}, we omit it from this work. 
Using these indices of refraction, Fig.~\ref{fig: bragg mode profiles} depicts the 
fundamental mode of the $N_1$ fiber at a non-resonant operating wavelength.  
Note the glass ring induces fine ripples in the longitudinal component of the mode profile; this effect will appear again in later sections.

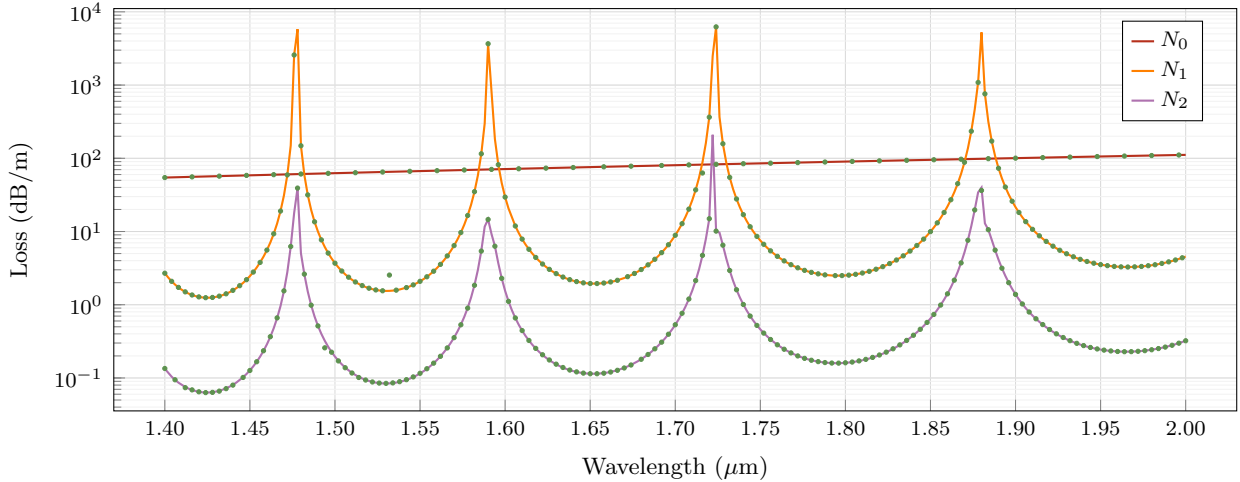
\begin{figure}
	\begin{tikzpicture}[scale=1]
	\colorlet{dotcolor}{OliveGreen!80}
	\newcommand{\dotsize}{.8pt}
	\newcommand{\linesize}{.8pt}
		\begin{axis}[wavelength,
			width=\textwidth,
		    height=.42\textwidth,
		    ymode=log,
		    enlargelimits=.05,
		    ]
			\addplot+[
			    color=BrickRed,
			    mark=*,
			    mark size=0pt,
			    line width=\linesize,
			    ]
			table[]{figures/data/bragg/N0/analytic.dat};
			\addlegendentry{$N_0$}
			\addplot+[
			    color=dotcolor,
			    mark=*,
			    mark size=\dotsize,
			    line width=0pt,
			    only marks,
			    mark options={fill=dotcolor},
			    forget plot,
			    ]
			table[]{figures/data/bragg/N0/numeric.dat};
			\addplot+[
			    color=orange,
			    mark=*,
			    mark size=0pt,
			    line width=\linesize,
			    ]
			table[]{figures/data/bragg/N1/analytic.dat};
			\addlegendentry{$N_1$}
			\addplot+[
			    color=dotcolor,
			    mark=*,
			    mark size=\dotsize,
			    line width=0pt,
			    only marks,
			    mark options={fill=dotcolor},
			    forget plot,
			    ]
			table[]{figures/data/bragg/N1/numeric.dat};
			\addplot+[
			    color=Orchid,
			    mark=*,
			    mark size=0pt,
			    line width=\linesize,
			    ]
			table[]{figures/data/bragg/N2/analytic.dat};
			\addlegendentry{$N_2$}
			\addplot+[
			    color=dotcolor,
			    mark=*,
			    mark size=\dotsize,
			    line width=0pt,
			    only marks,
			    mark options={fill=dotcolor},
			    forget plot,
			    ]
			table[]{figures/data/bragg/N2/numeric.dat};
		\end{axis}
	\end{tikzpicture}
\caption{Spectral loss profiles of Bragg fibers:  good agreement is observed 
between the analytical results (curves) and numerical results (dots).}
\label{fig: bragg loss}
\end{figure}

Figure~\ref{fig: bragg loss} shows the spectral loss profiles for several Bragg fibers
with different numbers of complete anti-resonant layers. The green dots show results of the numerical method,  showing
nearly perfect overlap with semi-analytic results for almost all values. 
The $N_0$ fiber, consisting of a low index core and high index cladding extending to the
computational boundary, does not show spikes in CL, but $N_1$ and $N_2$ fibers, both of 
which contain complete
anti-resonant layers, do. 
 It is possible to view the $N_0$ fiber as the result of 
modeling either the $N_1$ or $N_2$ fibers with the computational boundary placed in their 
first high index layer.  This choice of placement removes the loss spike behavior from their
profiles, hence Figure~\ref{fig: bragg loss} is the first demonstration that this modeling
choice can have a dramatic effect on the loss profile.  More such demonstrations appear in later sections.

The peaks in loss profiles occur at wavelengths $\lambda_m$ where standing waves occur in the
high index rings.  A good approximation for these wavelengths is given by
\begin{align}
\lambda_m &= \frac{2n_1d}{m}\big[ ( n_2 / n_1) ^2 - 1\big]^{1/2}, \quad m = 1,2, \ldots,
\label{eq: spikes}
\end{align}
which, though derived for the slab waveguide, gives good values for the
locations of the loss spikes for the Bragg fiber for wavelengths much smaller
than the core radius \cite{litchinitserAntiresonantReflectingPhotonic2002}.

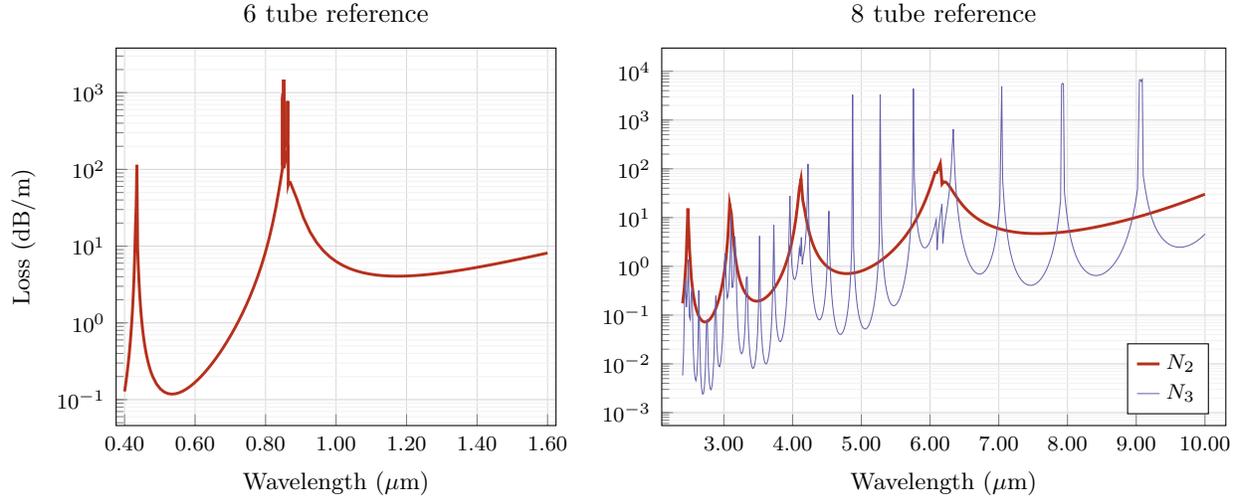
\begin{figure}
     \centering
		\begin{tikzpicture}
    \begin{groupplot}[ group style={group size=2 by 1, horizontal sep=40pt, ylabels at=edge left,  xlabels at=edge bottom},
        wavelength,
	    ymode=log,
		legend style={legend pos=south east, font=\tiny}
	]
    \nextgroupplot[title={6 tube reference},
            width=.45\textwidth,
       		height=0.4\textwidth,
       		enlarge x limits=.02,
       		] 
			\addplot+[
			    color=BrickRed,
			    mark size=0pt,
			    line width=1.1pt,
			    ]
			table[]{figures/data/bragg/6tube/analytic.dat};

    \nextgroupplot[title={8 tube reference},
           		enlarge x limits=.04,
            width=.55\textwidth,
        height=0.4\textwidth,]
			\addplot+[
			    color=BrickRed,
			    mark size=0pt,
			    line width=1.1pt,
			    ]
			table[]{figures/data/bragg/8tube/analytic.dat};
			\addlegendentry{$N_2$}
			\addplot+[
			    color=Blue!70,
			    mark size=0pt,
			    line width=.1pt,
			    ]
			table[]{figures/data/bragg/8tube/n3.dat};
			\addlegendentry{$N_3$}    \end{groupplot}
\end{tikzpicture}
		\caption{Loss profiles of reference Bragg fibers for the 6- and 8-tube ARFs}
\label{fig: ref bragg ARFs}
\end{figure}

The results above were for a general Bragg fiber with dimensions chosen
to clarify the structure of the loss
profiles and modes.  The more complex 6- and 8-tube anti-resonant fibers
chosen as examples for detailed study in later sections, have $N_2$ reference
Bragg fibers associated with them.  The thickness of the inner glass ring of
the model $N_2$ fiber is chosen to be the same thickness as the capillary
tube in the microstructure fiber.  The width of the subsequent air layer
is chosen based on the optimization process in
\cite{polettiOptimisingPerformancesHollow2011}, and corresponds approximately
to the diameter of the capillary tubes.  For 6-tube fiber, the reference core
radius is $15 \micrometer$, ring thickness is $0.42 \micrometer$ and
air layer width $12.48 \micrometer$.  For the 8-tube fiber, these measurements
are $59.5 \micrometer, 6\micrometer,$ and $25.5 \micrometer$ respectively.
Their loss profiles are shown in the blue curves of Figure~\ref{fig: ref bragg ARFs}.

Being of the $N_2$ type, these reference fibers allow the cladding to 
extend to infinity.  If we truncate the cladding in these
reference fibers, i.e., use the $N_3$ configuration, further loss spikes appear.
The inclusion of further materials in the outer regions of the computational
domain changes the fiber loss profile.
For the 8-tube fiber, we used the published scanning electron micrograph
image \cite[Fig.~3]{kolyadinLightTransmissionNegative2013} to estimate the
thickness of the outer glass cladding, and we included this
to make an $N_3$ reference fiber.  The spectral loss profile of this fiber
is shown in light blue in Figure~\ref{fig: ref bragg ARFs}.
These extra resonant peaks correspond to values predicted by
\eqref{eq: spikes} when $d$ is the thickness of the cladding. 


In this section we have seen two examples of the effects of changing the material
configuration at the boundary of the computational domain, and verified that our
numerical method can capture the results of these changes.  In both cases, if 
the boundary is placed in what would otherwise be a complete anti-resonant layer,
a set of loss spikes is eliminated.  
In the next section, we will see that loss profiles for ARFs behave in a similar fashion.

\section{ARF loss profiles in two outer configurations}
\label{sec:arf-CL}

In this section, we study the loss profiles for several ARF fibers using different
material configurations at the computational boundary. We have already seen 
in Section~\ref{sec:bragg-fiber} that the
number of resonant peaks in the loss profiles of the reference Bragg fibers for these fibers
depends on how the outer layer is modeled.
Such differences will become more pronounced
in a full-featured numerical model of the ARF.

\begin{figure}
     \centering
     \begin{subfigure}[b]{0.37\textwidth}
         \centering
         \includegraphics[width=.9\textwidth]{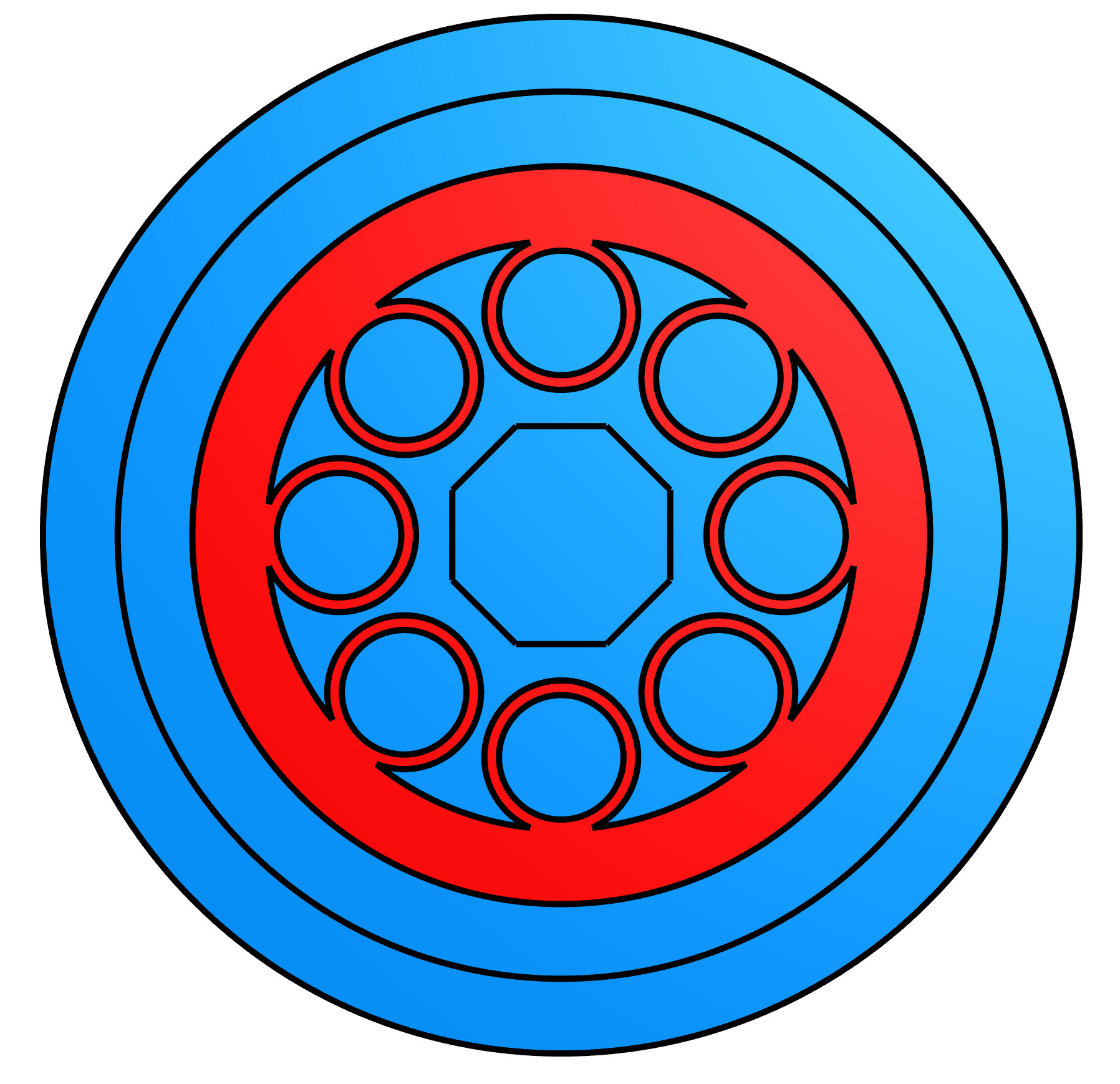}
         \caption{Geometry.}
         \label{fig:kolyadin-geom}
     \end{subfigure}
     \hspace{.32in}
     \begin{subfigure}[b]{0.37\textwidth}
         \centering
         \includegraphics[width=.9\textwidth]{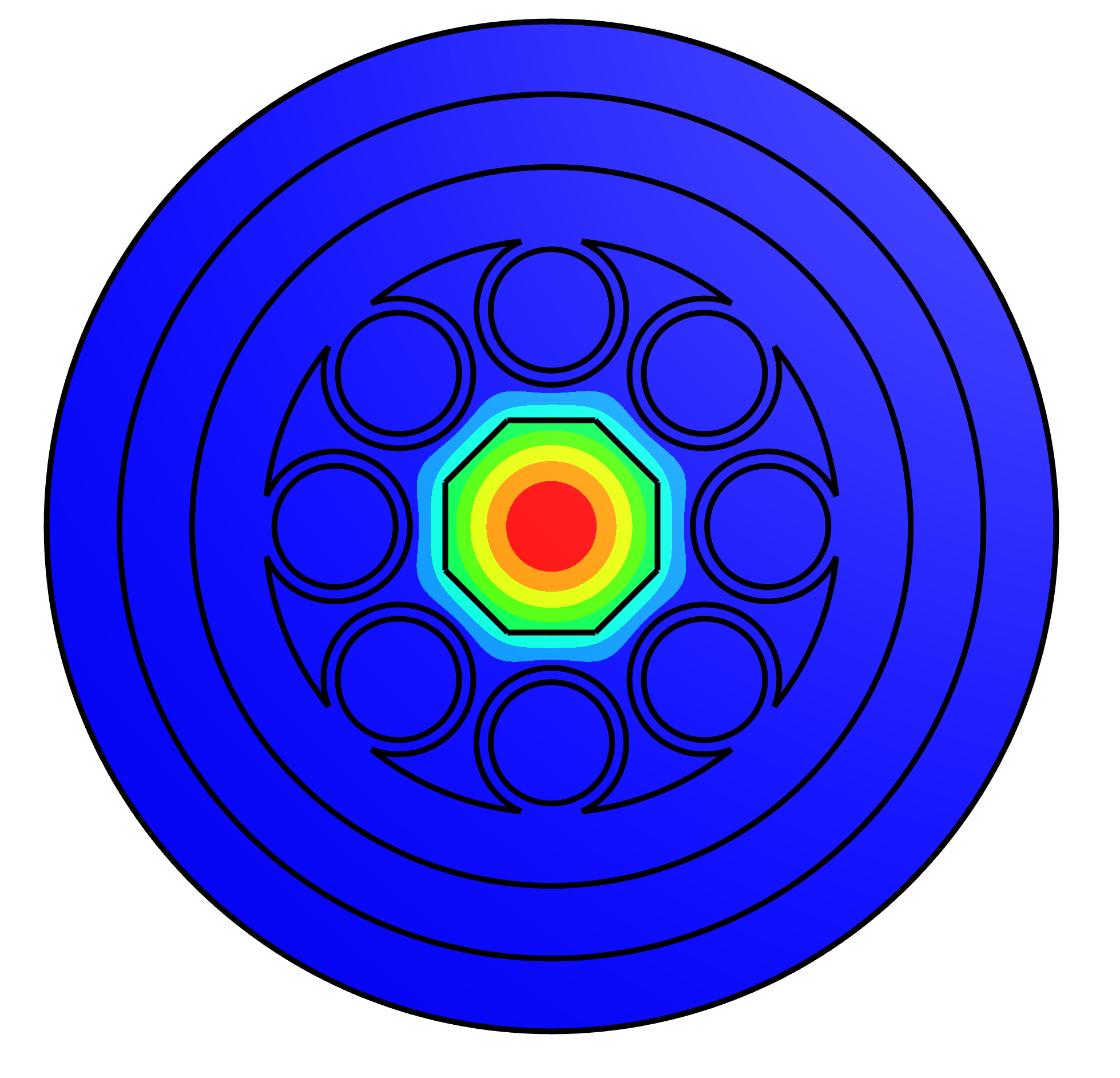}
         \caption{Transverse field magnitude.}
         \label{fig:Kolyadin-fm}
      \end{subfigure}
 \caption{Geometry (a) and fundamental mode profile (b) of 8-tube anti-resonant hollow-core fiber.  In the geometry, red indicates glass, blue indicates air, and black curves demarcate distinct computational regions (used for numerical purposes).  Note that this shows the fiber in the $C_1$ configuration.  
 }
\label{fig: kolyadin}
\end{figure}

For ease of reference, we define a notation for these different
material configurations.  The microstructures of anti-resonant fibers are
supported inside one or more dielectric layers
such as a glass cladding and/or polymer coatings.
The fiber is said to be in {\em Configuration~1} or $C_1$, if the outermost
dielectric layer
is allowed to terminate and the model is placed in a
lower index ambient material
(such as air).  
If the model instead extends the outermost high index
layer to the end of the computational
domain, then we say it is in {\em Configuration~0} or $C_0$.
Figure~\ref{fig:kolyadin-geom} shows the geometry of the 
8-tubed ARF of~\cite{kolyadinLightTransmissionNegative2013} in the $C_1$
configuration.

For numerical computations, we use 
the perfectly matched later (PML) to terminate the
outer infinite layer (made of air in the $C_1$ model and of glass in the $C_0$ model).
In  Figure~\ref{fig:kolyadin-geom},
the end of the computational domain is marked by the  outermost
circle. The next smaller circle there, 
drawn inside of the air-domain, marks the beginning of the PML region.
The octagonal inner region is for computational purposes only (such as mesh size control). All
regions colored blue in the figure represent air.

\begin{table}
%


    \centering
    \begin{tabular}{c g c g c g c g c g}
 		\rowcolor{white} &  & \multicolumn{4}{c}{Core Parameters ($\mu \mbox{m}$)} \
 		&  \multicolumn{4}{c}{Configuration Parameters ($\mu \mbox{m}$)} \\
        \cmidrule(lr){3 - 6}  \cmidrule(lr){7 - 10}
        \rowcolor{white} Fiber & $N_{\mbox{\tiny cap}}$ & $R_{\mbox{\tiny core}}$ & $R_{\mbox{\tiny cap}}$ \
        								       &  $t_{\mbox{\tiny cap}}$  & $e_{\mbox{\tiny cap}}$   & $t_{\mbox{\tiny clad}}$ \
        									   & $t_{\mbox{\tiny poly}}$ & $t_{\mbox{\tiny buffer}}$ & $t_{\mbox{\tiny pml}}$\\[.01in]
        \midrule
        6-tube &  6  &  15.0  &  12.48  &  0.42  &  .025  &  10.0  &  10.0   &  10.0  &  30.0  \\[.05in]
        8-tube &  8  &  59.5  &  25.50  &  6.00  &  2.00  &  30.6  &  15.3  &  15.3  &  22.95   \\[.01in]
        \bottomrule 
    \end{tabular}
\vspace{.1in}
\caption{ARF Parameters}
\label{tab:params}
\end{table}

The geometry of the microstructured core region of the fibers
considered here can be determined by 5 parameters:\footnote{It is also possible
to use the azimuthal separation $d$ of capillary tubes as done in \cite{polettiNestedAntiresonantNodeless2014}, in which
case the $N_{\mbox{\tiny cap}}$ is determined by this.}
 the number of capillary
tubes $N_{\mbox{\tiny cap}}$, the core radius 
$R_{\mbox{\tiny core}}$, the inner capillary radius $R_{\mbox{\tiny cap}}$,
the capillary thickness $t_{\mbox{\tiny cap}}$ and the embedding distance
$e_{\mbox{\tiny cap}}$.
Thicknesses of the further layers then determine the rest of the model.
For the 6- and 8-tube fibers, we used the parameters shown in
Table~\ref{tab:params}.

\begin{figure}
	\centering
			\begin{tikzpicture}
		\begin{axis}[wavelength,
		    ymode=log,
		    enlargelimits=.05,
		    width=\textwidth,
		    height=.4\textwidth,
			legend style={legend pos=south east, font=\tiny},
		    ]
			\addplot+[
			    color=blue!75,
			    mark=*,
			    mark size=0pt,
			    line width=.001pt,
			    ]
			table[]{figures/data/arf/8tube/N3config_ref0.dat};
			\addlegendentry{$C_1$}
			\addplot+[
			    color=BrickRed,
			    mark=*,
			    mark size=0pt,
			    line width=.75pt,
			    ]
			table[]{figures/data/arf/8tube/N2config.dat};
			\addlegendentry{$C_0$}
		\end{axis}
	\end{tikzpicture}
		\caption{Spectral loss profile of 8-tube ARF}
                \label{fig:kolyadin-spectral-loss-profile}
\end{figure}
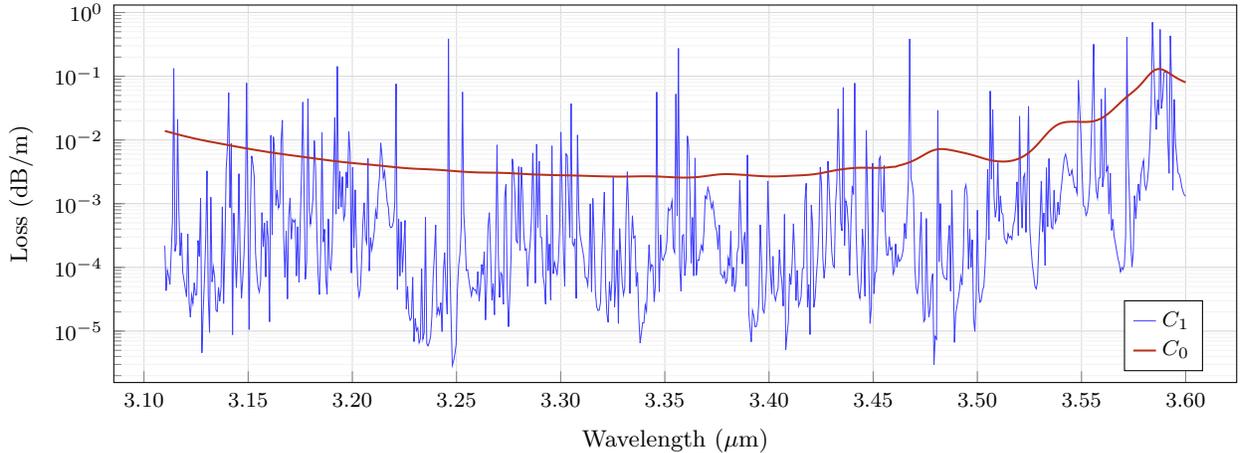
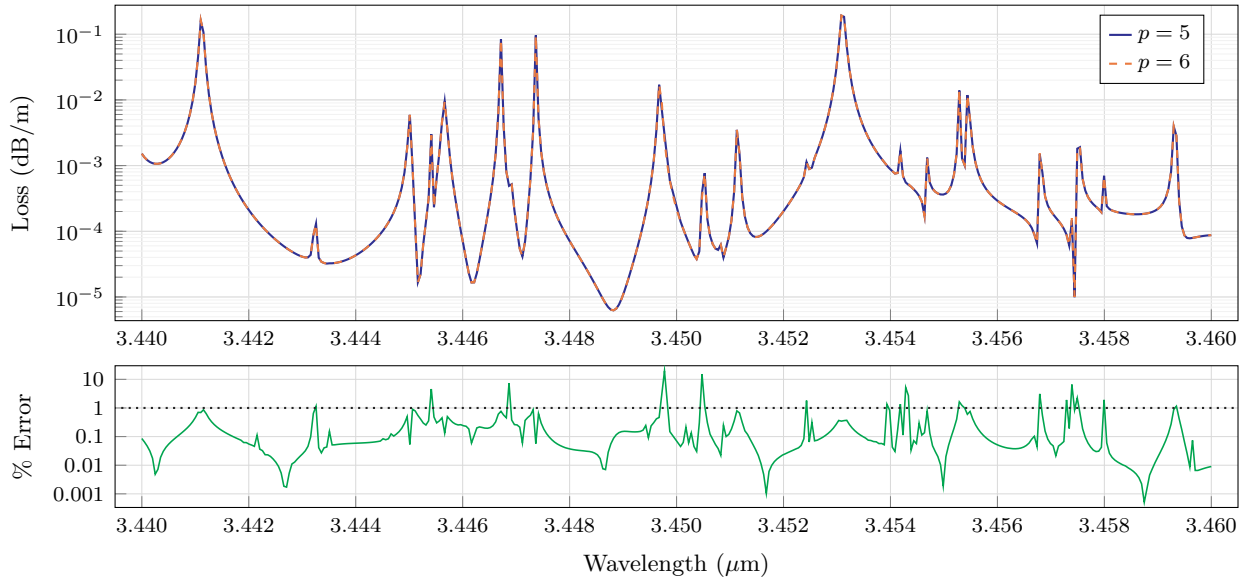
\begin{figure}
	\centering
				\begin{tikzpicture}
    \begin{groupplot}[
        group style={
            group size=1 by 2,
            vertical sep=.6cm,
            xlabels at=edge bottom,
        },
        wavelength,
        enlarge x limits=0.025,
	    ymode=log,
		x tick label style={font=\tiny, /pgf/number format/zerofill,  /pgf/number format/precision=3},
	]
    \nextgroupplot[width=\textwidth,
    						 height=.35\textwidth,
	    					 enlarge y limits=.035,
    						 ]
			\addplot+[
			    color=Blue,
			    enlargelimits=0.0,
			    mark=*,
			    mark options={fill=orange!80!blue,},
			    mark size=0pt,
			    line width=.8pt,
			    ]
			table[]{figures/data/arf/8tube/ref0_p6_subint.dat};
			\draw[very thick] (0,0) rectangle (.5,.5);
			\addlegendentry{$p=5$}
			\addplot+[
			    color=Orange!95!Blue,
			    enlargelimits=0.0,
			    mark=*,
			    mark options={fill=Blue!40},
			    mark size=0pt,
			    line width=.8pt,
			    dashed,
			    ]
			table[]{figures/data/arf/8tube/ref0_p5_subint.dat};
			\addlegendentry{$p=6$}
    \nextgroupplot[width=\textwidth,
    						 height=.21\textwidth,
    						 xmin=3.44,
    						 xmax=3.46,
    						 ylabel={\% Error},
	    					 log ticks with fixed point,
	    					 ytick={10, 1, .1, .01, .001},
	    					 enlarge y limits=.04,
   					  		 ]
			\addplot+[
			    color=Green,
			    mark=*,
			    mark options={fill=orange!90!blue},
			    mark size=0pt,
			    line width=.6pt,
			    ]
			table[]{figures/data/arf/8tube/ref0_p6p5_rel_error.dat};
			\addplot +[mark=none, color=black!85, dotted, 
							  line width=.8pt] coordinates {(3.4392, 1) (3.4608, 1)};    
		\end{groupplot}
	\end{tikzpicture}
		\caption{Spectral loss profile sub-interval for 8-tube ARF showing numerical stability of CL at each wavelength point.}
                \label{fig:residual}
\end{figure}

Figure~\ref{fig:kolyadin-spectral-loss-profile} shows the loss profile
of the 8-tube fiber in configurations $C_0$ and $C_1$.  The
differences in the computed CL values of $C_0$ and $C_1$ models are
dramatic. The $C_1$ loss profile has profound variation, on a scale 
difficult to resolve even with the 800 samples taken over this wavelength 
interval. 
The total variation seen is nearly 6 orders of magnitude.
In stark contrast, the $C_0$ profile has little variation on a granular
level and total variation of about one order of magnitude. 

The apparently chaotic oscillations in the $C_1$ curve are
not due to numerical errors. 
Each point in the plot is obtained with enough numerical degrees of freedom that the computed CL values do not vary significantly under further mesh refinement or under variations in PML parameters. 
As an example to demonstrate this, we have included a zoomed-in view of the plot in
Figure~\ref{fig:kolyadin-spectral-loss-profile} over a smaller
subinterval of wavelengths, in Figure~\ref{fig:residual}. The second
plot of the figure shows that under one further increase of element
order, the maximal change in computed CL value is less than 1\% for
almost all values.  Locations in the subinterval where
error exceeded 1\% were typically associated with loss spikes.
While more
investigation is needed into convergence of values at these locations,
it is clear that the visual appearance the loss profiles will remain
unchanged.

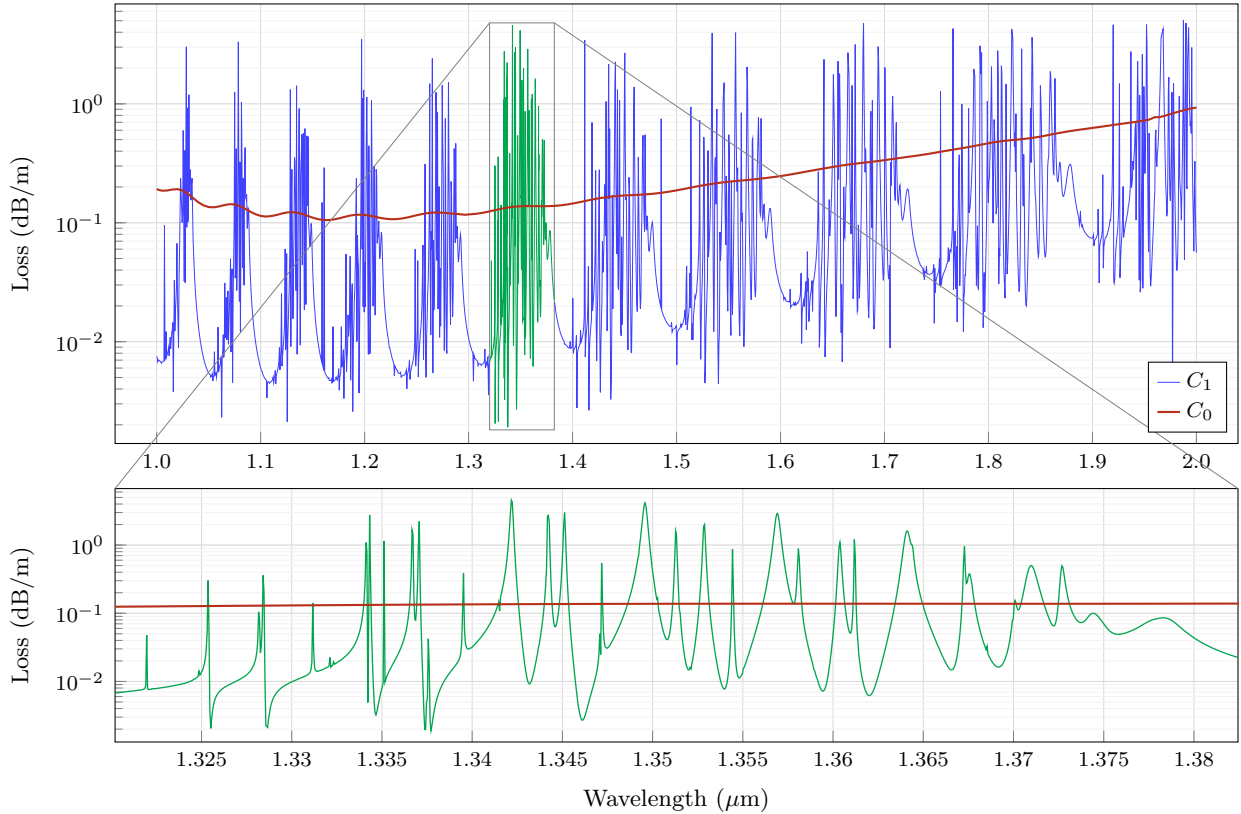
\begin{figure}[b]
\centering
\pgfplotstableread{figures/data/arf/6tube/C1_subinterval.dat}\Czerosub

\SetElement{\Czerosub}{0}{0}{\xmin}
\pgfplotstablegetrowsof{\Czerosub}
\pgfplotstablegetelem{\the\numexpr\pgfplotsretval-1}{0}\of{\Czerosub} \let\xmax\pgfplotsretval

\newcommand{\findYlims}[1]{%
\pgfplotstablegetrowsof{#1}
	\pgfmathtruncatemacro{\LastRowNo}{\pgfplotsretval-1}
\pgfplotstablesort[sort key={1}]{\sorted}{#1}%
    \pgfplotstablegetelem{\LastRowNo}{1}\of{\sorted}%
\xdef\Ymax{\pgfplotsretval}%
    \pgfplotstablegetelem{0}{1}\of{\sorted}%
\xdef\Ymin{\pgfplotsretval}}

\findYlims{\Czerosub}

\newcommand{\yscaling}{.05}

\begin{tikzpicture}
    \begin{groupplot}[
        group style={
            group size=1 by 2,
            vertical sep=.6cm,
            xlabels at=edge bottom,
        },
        wavelength,
	    ymode=log,
		legend style={legend pos=south east, font=\tiny},
	]
    \nextgroupplot[width=\textwidth,
    						 height=.45\textwidth,
        					 enlarge x limits=0.04,
	    					 enlarge y limits=0.04,
	   						 legend to name=grouplegend,
			x tick label style={font=\tiny, /pgf/number format/precision=1},
    						 ]
\draw[very thin, black!50] (axis cs:\xmin, \Ymin-\yscaling*\Ymin) rectangle (axis cs:\xmax, \Ymax + \yscaling*\Ymax);
		\coordinate (UL) at (axis cs:\xmin,\Ymax + \yscaling*\Ymax);
		\coordinate (UR) at (axis cs:\xmax,\Ymax + \yscaling*\Ymax);
			\addplot+[
			    color=blue!75,
			    enlargelimits=0.0,
			    mark=*,
			    mark options={fill=orange!80!blue,},
			    mark size=0pt,
			    line width=.1pt,
			    ]
			table[]{figures/data/arf/6tube/C1L.dat};
			\addlegendentry{$C_1$}
			\addplot+[
			    color=Green,
	    		enlarge y limits=.05,
			    mark=*,
			    mark options={fill=Blue!40},
			    mark size=0pt,
			    line width=.1pt,
			    forget plot,
			    ]
			table[]{figures/data/arf/6tube/C1_subinterval.dat};
%
			\addplot+[
			    color=blue!75,
			    enlargelimits=0.0,
			    mark=*,
			    mark options={fill=orange!80!blue,},
			    mark size=0pt,
			    line width=.1pt,
			    forget plot,
			    ]
			table[]{figures/data/arf/6tube/C1R.dat};
%
			\addplot+[
			    color=BrickRed,
			    enlargelimits=0.0,
			    mark=*,
			    mark options={fill=orange!80!blue,},
			    mark size=0pt,
			    line width=.8pt,
			    ]
			table[]{figures/data/arf/6tube/C0.dat};
			\addlegendentry{$C_0$}
    \nextgroupplot[width=\textwidth,
    						 height=.3\textwidth,
	    					 enlarge y limits=\yscaling,
	    					 enlarge x limits=0,
						     xmin=\xmin,
						     xmax=\xmax,
		x tick label style={font=\tiny, /pgf/number format/zerofill=false,  /pgf/number format/precision=3},
			       			 ]
			\addplot+[
			    color=Green,
			    mark=*,
			    mark options={fill=orange!90!blue},
			    mark size=0pt,
			    line width=.5pt,
			    ]
			table[]{figures/data/arf/6tube/C1_subinterval.dat};
			\addplot+[
			    color=BrickRed,
			    mark=*,
			    mark options={fill=orange!90!blue},
			    mark size=0pt,
			    line width=.8pt,
			    ]
			table[]{figures/data/arf/6tube/C0_subinterval.dat};
		\end{groupplot}
\draw[very thin, black!50] 
	(UL) -- (group c1r2.north west) node {};
\draw[very thin, black!50] 
	(UR) -- (group c1r2.north east) node {};
	\node[anchor=south east] at (group c1r1.south east){\pgfplotslegendfromname{grouplegend}};   
\end{tikzpicture}
\caption{Spectral loss profile of 6-tube ARF.}
\label{fig:6tube-loss-profile}
\end{figure}

We repeated this study for the 6-tube ARF, again using $C_0$ and $C_1$ models.
The  results
are shown in Figure~\ref{fig:6tube-loss-profile}.
The message is the same: the CL values from $C_0$ and $C_1$ models are
vastly different.
In comparison to the results for the
8-tube fiber, more structure is now apparent  for the spectral
loss profile associated to the $C_1$ model.
As in that fiber, there is a
great deal of variation, but here there are regions of intense variation, 
interspersed with regions of relative smoothness.  Notably, 
the highly variable portions in $C_1$  configuration occur in the same
regions as subtle waves in $C_0$ configuration.  This is
especially visible on the left portion of the plot.  This suggests that
the localization of high variability is somehow inherent in the design of the 
core and microstructured cladding regions (since
it is that which remains unchanged in the two configurations).
The lower portion of Figure~\ref{fig:6tube-loss-profile} shows an inset
(note the scale change) focused on one of the high variance regions.  Taking enough wavelength samples for high resolution, we
can see that the profile does indeed have a discernible structure in these
areas, with clear resonant peaks and anti-resonant troughs. 

Therefore, this study has illustrated that the outer layer material configurations have a more pronounced effect on the ARF fibers' loss profiles than they did for the Bragg fiber designs. 
The levels of CL variation computed in the model are greater than those observed experimentally.  
In the next section
we analyze how the inclusion of a layer of material loss may affect these results, bringing the modeled loss profiles more in line with observation.

\section{Inclusion of Lossy Polymer}
\label{sec:polymer}

The vast difference between the previously seen loss profiles of $C_0$ and $C_1$ models of  ARF fibers naturally prompts this question: which
model approximates reality best? The infinite
cladding of $C_0$ is obviously not realizable in practice, and their loss profiles
are smoother than experimentally observed results (see \cite{hayesAntiresonantHollowCore2017}).  Similarly,
the $C_1$ model with air directly surrounding the first
layer of glass cladding does not represent a fiber in any
real life scenario, since all optical fibers have coatings that are typically
many times thicker than their core diameters, and the resulting loss
profiles have correspondingly unrealistic levels of total variation.  In this section,
we attempt to refine our analysis by including these coating layers into our model.
These layers are applied first to the Bragg fiber design to gain insight where semi-analytic
solutions are available and to show that our numerical method still recovers the correct loss values. 
Afterwards, this lossy material layer is applied to the ARF fiber designs.

In \cite[p.~63]{marcuseLightTransmissionOptics1982}, it is noted that  ``in
most practical cases cladding modes will be suppressed by a lossy coating on
the outside of the fiber.''  
Such lossy coatings are present on any real-life fiber to provide mechanical
protection and structural support.  Even small distortions to an uncoated fiber
can cause significant loss, hence a good deal of research has been done on
optimizing optical jackets for the minimization of transmission loss, see
\cite{glogeOpticalfiberPackagingIts1975, shiueDesignDoublecoatedOptical1992,
cocchiniLateralRigidityDoublecoated1995}.  This suggests
that the inclusion of these coating layers in the model may lead to reduced loss,
possibly decreasing the unrealistic total variation of fiber in the $C_1$ configuration.

\begin{figure}[b]
	\centering
			\begin{tikzpicture}
		\begin{axis}[wavelength,
		    ymode=log,
		    enlargelimits=.05,
		    height=.35\textwidth,
		    width=.85\textwidth,
		    ylabel={$k$},
			legend style={legend pos=south east, font=\tiny},
		    ]
			\addplot+[
			    color=blue,
			    mark=*,
			    mark size=0pt,
			    line width=.4pt,
			    ]
			table[]{figures/data/refidx/PVC_k.dat};
			\addlegendentry{PVC}
			\addplot+[
			    color=BrickRed,
			    mark=*,
			    mark size=0pt,
			    line width=.4pt,
			    ]
			table[]{figures/data/refidx/PMMA_k.dat};
			\addlegendentry{PMMA}
		\end{axis}
	\end{tikzpicture}
		\caption{Extinction coefficients of some common polymers}
	\label{fig:extcoeffs}
\end{figure}
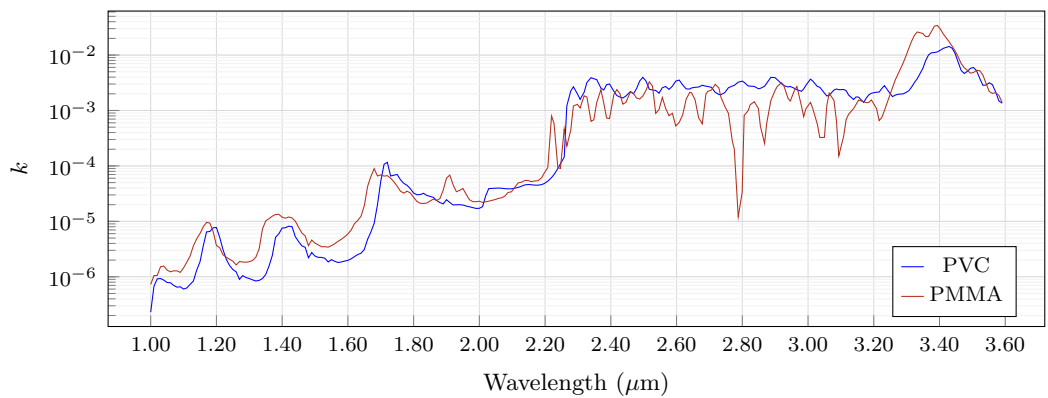

Material losses can be accounted for with the addition of a complex component
to the refractive index \cite{snyderOpticalWaveguideTheory2024} called the
{\em extinction coefficient}.  Information on such components in general is limited,
and often fiber coatings are proprietary.  Figure~\ref{fig:extcoeffs} shows
extinction coefficient values for poly vinyl-chloride (PVC) taken from 
poly(methyl methacrylate) (PMMA) \cite{zhangComplexRefractiveIndices2020} and
\cite{polyanskiyRefractiveIndexInfo}.  We can see that the material loss is highly
dependent on wavelength.

For the purpose of our current studies, we utilize values for the extinction coefficient ($k$) and
polymer thickness that best display the effects under consideration.  Polymer thickness
is typically on the order of the core radius, and values for $k$ range from $10^{-3}$
to $10^{-1}$.  When including polymer layers in real life fiber analysis, it would be 
necessary to balance increased computational load with the level of desired detail
sought in the loss profile.  Our goal here is to draw attention to the possibility
that more information may be found in loss profiles by including further materials
in the computational model.  We include recommendations for best practices for 
this in the discussion.

\subsection{Bragg fiber model with polymer}
\label{subsec:bragg-poly}

We begin by extending the Bragg model to include 
a lossy polymer layer.  The Bragg fiber still has semi-analytic 
solutions in this setting.  This allows us
to quickly explore the ramifications of including this layer in our model
and to verify that our numerical method can still capture
the loss values in this new case.  

\begin{figure}
     \centering
     \begin{subfigure}[b]{0.32\textwidth}
         \centering
         \includegraphics[width=\textwidth]{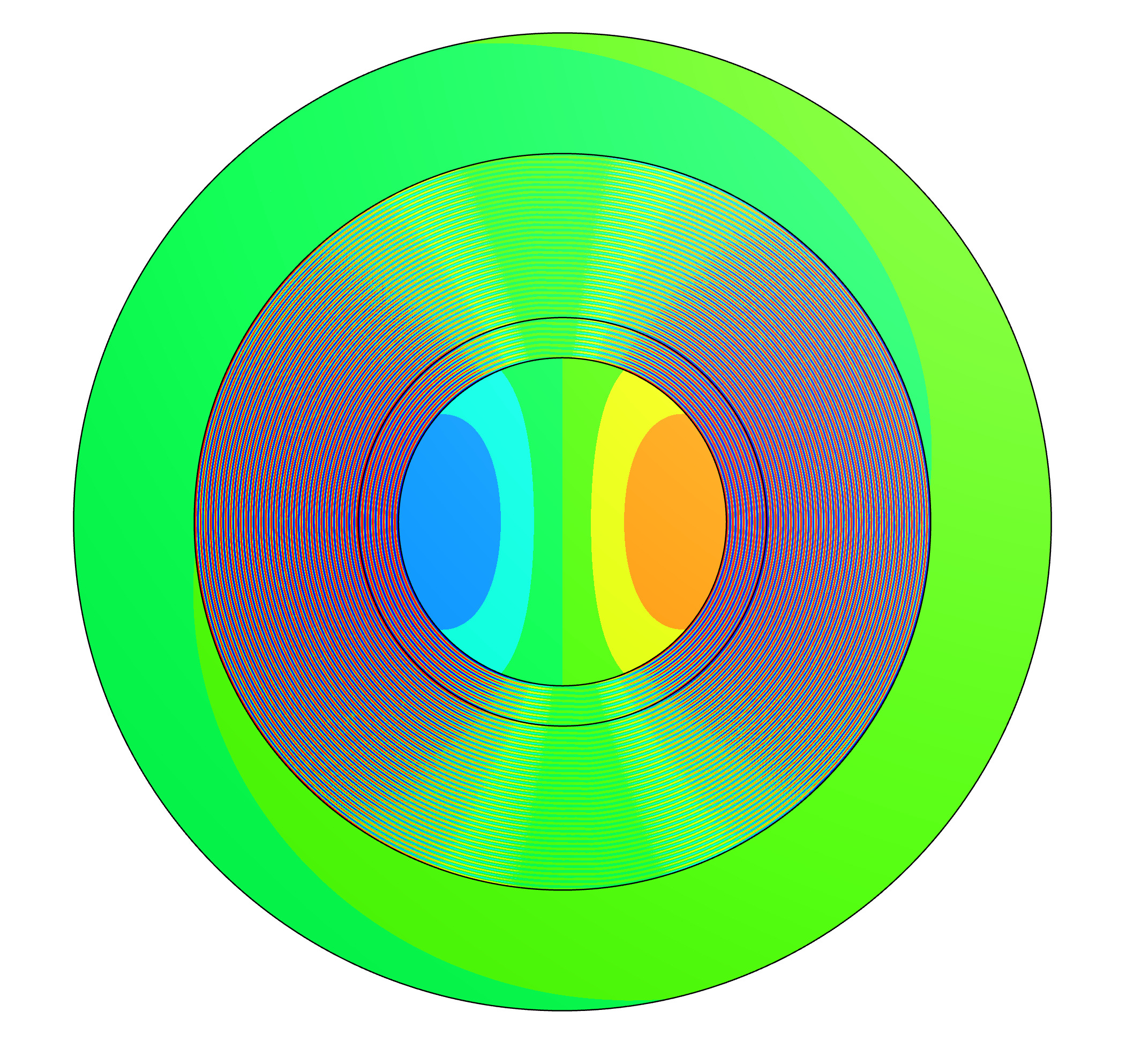}
         \caption{No material loss}
         \label{fig:mode-bragg-poly-lossless}
     \end{subfigure}
     \hspace{.25in}
     \begin{subfigure}[b]{0.32\textwidth}
         \centering
         \includegraphics[width=\textwidth]{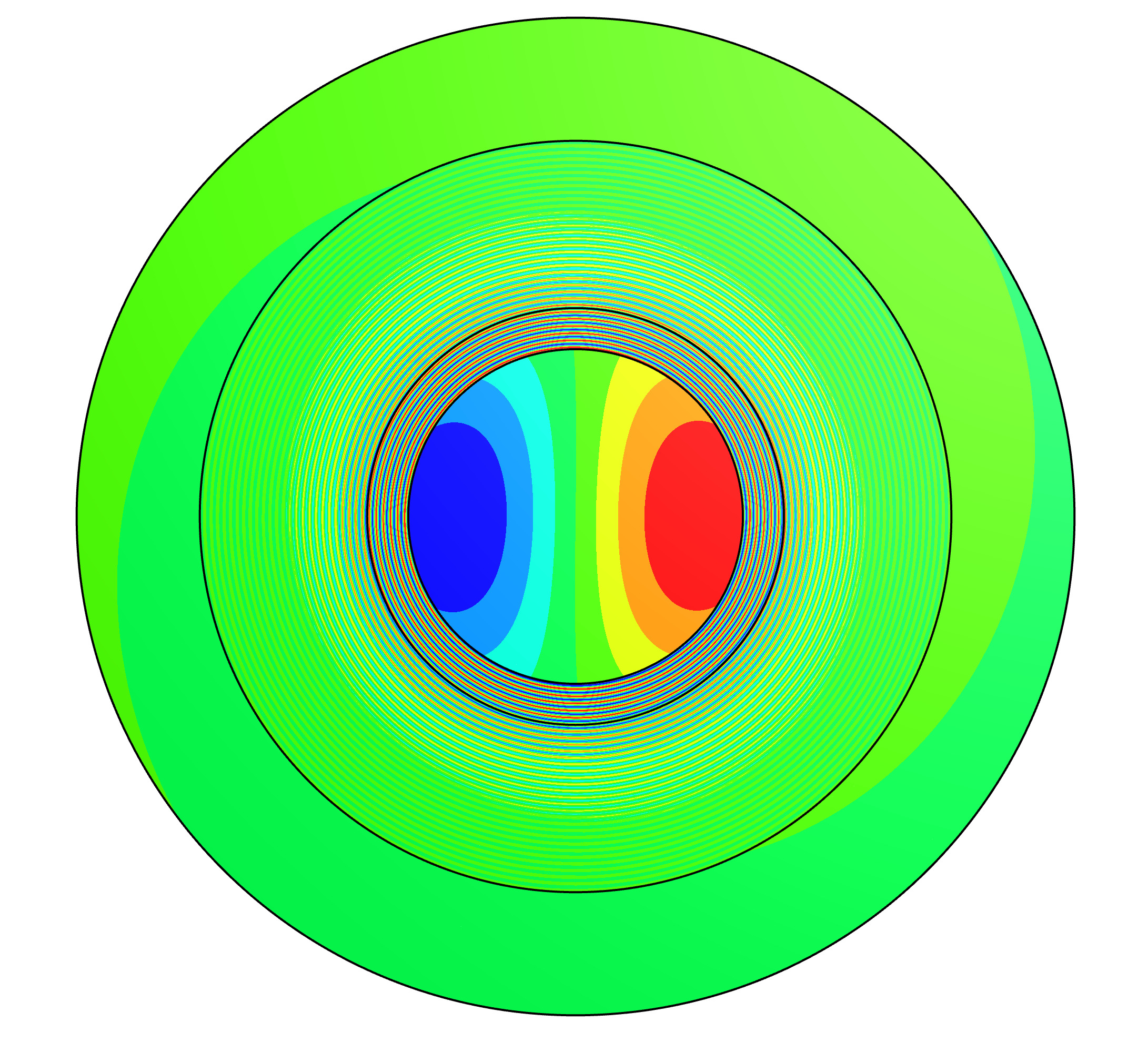}
         \caption{With material loss, $k$ = 0.01}
         \label{fig:mode-bragg-poly-lossy}
     \end{subfigure}
 \caption{Bragg fiber with polymer: Fundamental mode longitudinal components.}
\label{fig:braggpolymodes}
\end{figure}

\begin{figure}[b]
     \centering
			\begin{tikzpicture}
		\colorlet{dotcolor}{OliveGreen!90}
		\newcommand{\dotsize}{.8pt}
		\newcommand{\linesize}{.8pt}
		\begin{axis}[wavelength,
		    ymode=log,
		    height=.4\textwidth,
		    width=.9\textwidth,
		    enlargelimits=.05,
		    legend pos=outer north east,
		    ]
			\addplot+[
			    color=BrickRed,
			    mark=*,
			    mark size=0pt,
			    line width=1.1pt,
			    ]
			table[]{figures/data/bragg/poly/k_inf.dat};
			\addlegendentry{$N_0$ Bragg};	
			\addplot+[
			    color=Gray,
			    mark=*,
			    mark size=0pt,
			    line width=\linesize,
			    ]
			table[]{figures/data/bragg/poly/k_0.01.dat};
			\addlegendentry{$k = 0.01$};
			\addplot+[
			    color=Magenta,
			    mark=*,
			    mark size=0pt,
			    line width=\linesize,
			    ]
			table[]{figures/data/bragg/poly/k_0.005.dat};
			\addlegendentry{$k = 0.005$};
			\addplot+[
			    color=Cyan,
			    mark=*,
			    mark size=0pt,
			    line width=\linesize,
			    ]
			table[]{figures/data/bragg/poly/k_0.002.dat};
			\addlegendentry{$k = 0.002$};
			\addplot+[
			    color=dotcolor,
			    mark=*,
			    mark size=\dotsize,
			    line width=0pt,
			    mark options={fill=dotcolor},
			    forget plot,
			    ]
			table[]{figures/data/bragg/poly/k_002_numeric.dat};
			\addplot+[
			    color=Orange!90!Blue,
			    mark=*,
			    mark size=0pt,
			    line width=\linesize,
			    solid,
			    ]
			table[]{figures/data/bragg/poly/k_0.dat};
			\addlegendentry{$k = 0.00$};
		\end{axis}
	\end{tikzpicture}
 \caption{Polymer coated Bragg fiber spectral loss profile.  Green dots show results of numerical method.}
 \label{fig:bragg-poly-loss-profile}
\end{figure}
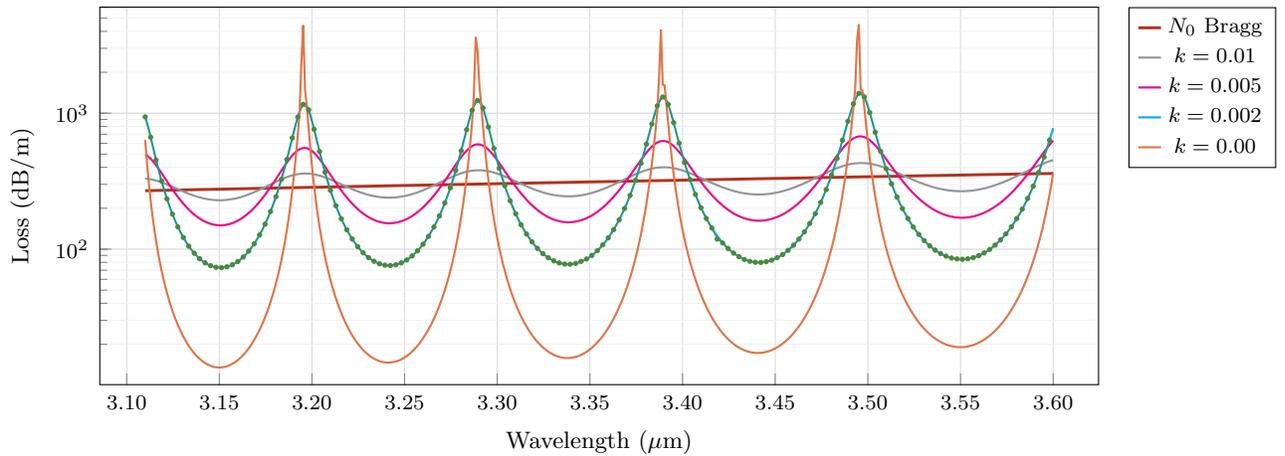

First, consider the $N_1$ fiber design with an added lossy polymer layer directly outside the glass ring 
and before the low index ambient material.  We set the
thickness of the polymer equal to the core radius, while keeping all 
remaining dimensions as
in Section~\ref{sec:bragg-fiber}. We fix the real part of the
refractive index of the polymer to be slightly higher than glass and
varied its imaginary component, i.e., $n = 1.5 + k \ii$ for varying
$k$-values.  (Here $\ii$ is the imaginary unit.) 

Plots~\ref{fig:mode-bragg-poly-lossless}
and~\ref{fig:mode-bragg-poly-lossy} depict the differences between the mode profile components for the polymer layer with and without material loss.
The first graphic in Figure~\ref{fig:braggpolymodes}
shows that when no material loss is included, the polymer
region acts in a similar way to the glass region in that it supports the fine ripples (reflections/standing waves) in the mode profile all the way to the air-polymer interface.  Thus, we expect the spectral loss
profile of this fiber to have spikes similar to previous $N_1$ models
(though in different locations).

Indeed, Figure~\ref{fig:bragg-poly-loss-profile} confirms this fact, while additionally illustrating the loss profiles for this fiber for a variety of polymer loss coefficients~$k$. 
As we expected, when $k=0$, the fiber has resonant loss spikes. 
The computational results demonstrate that larger polymer loss values~$k$ dampen the magnitude of these resonant loss spikes. 
Interestingly, the limiting case for loss profile seems to be that
of the $N_0$ Bragg fiber of the same core radius. It appears that increasing
material loss in the polymer makes both the glass annulus and its surrounding polymer layer progressively less resonant, until the loss profile becomes smooth. 
In this case, the glass ring, although terminating in the geometry, shows the same profile as if it extended to infinity. 


\subsection{ARF with Polymer}

With this intuition gathered from the Bragg fiber augmented with a lossy polymer coating, we repeat our previous CL computations for the 8-tube ARF, now likewise encompassing the glass cladding with a lossy polymer material. 
As with the Bragg fiber of the previous section, the polymer layer is followed by a low index material; a $C_1$-type configuration. 
The polymer thickness is set to $15.3 \micrometer$, as is the width of the buffer layer of air between the polymer and computational boundary layer. 
The PML layer is $22.95 \micrometer$ thick (see Table~\ref{tab:params}).

\begin{figure}[b]
     \centering
			\begin{tikzpicture}
		\begin{axis}[wavelength,
		    ymode=log,
		    width=.9\textwidth,
		    height=.4\textwidth,
		    enlargelimits=.025,
		    legend pos=outer north east,
		    reverse legend=true,
		    ]
			\addplot+[
			    color=Gray!30,
			    mark=*,
			    mark size=0pt,
			    line width=.02pt,
			    mark options={fill=OliveGreen!80},
			    ]
			table[]{figures/data/arf/8tube/k_0.0001.dat};
			\addlegendentry{$k = .0001$};
			\addplot+[
			    color=Periwinkle,
			    mark=*,
			    mark size=0pt,
			    line width=.4pt,
			    ]
			table[]{figures/data/arf/8tube/k_0.001.dat};
			\addlegendentry{$k = .001$};
			\addplot+[
			    color=Blue!80,
			    mark=*,
			    mark size=0pt,
			    line width=1pt,
			    mark options={fill=OliveGreen!80},
			    ]
			table[]{figures/data/arf/8tube/k_0.01.dat};
			\addlegendentry{$k = .01$};
			\addplot+[
			    color=BrickRed,
			    mark=*,
			    mark size=0pt,
			    line width=1.2pt,
			    ]
			table[]{figures/data/arf/8tube/k_inf.dat};
			\addlegendentry{$C_0$};
			\addplot+[
			    color=LimeGreen!80,
			    mark=*,
			    mark size=0pt,
			    line width=1.1pt,
			    dashed,
			    ]
			table[]{figures/data/arf/8tube/k_0.1.dat};
			\addlegendentry{$k=.1$};	
		\end{axis}
	\end{tikzpicture}
 \caption{Polymer coated 8-tube ARF spectral loss profile.}
\label{fig: kolyadin poly}
\end{figure}
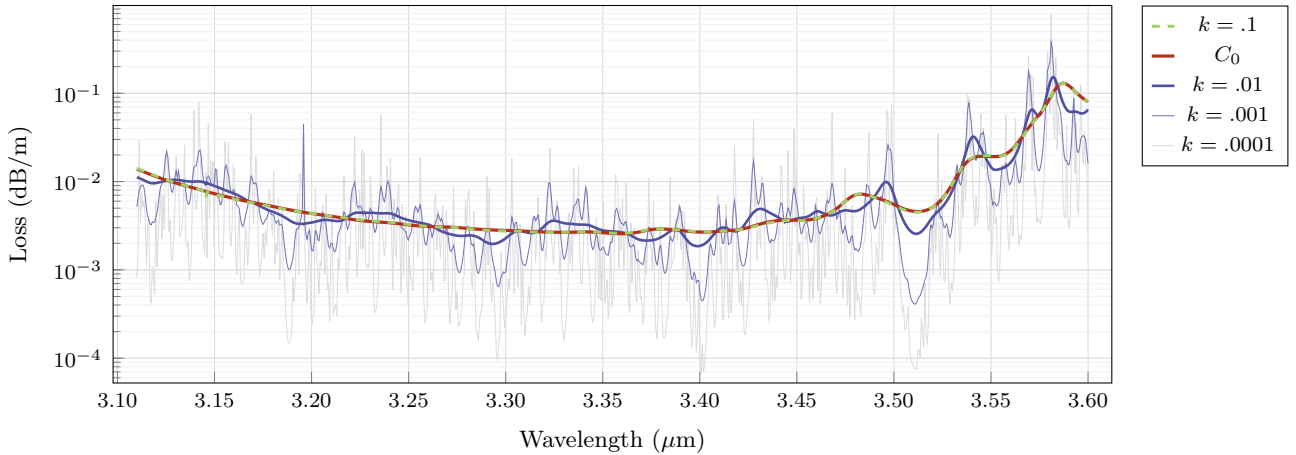

The computed 8-tube ARF spectral CL profile results are plotted in Fig.~\ref{fig: kolyadin poly} for polymer refractive indices of $1.5 + \ii k$, where $k = 0.1, 0.01, 0.001, 0.0001$.  Also pictured is the loss profile for ARF model in the $C_0$ configuration, which seems to be the limiting case for high values of $k$. 
As $k$ decreases, the variation in the spectral profile increases, approaching the
level of variation seen in the prior $C_1$ model results (without the polymer layer), as seen in Figure~\ref{fig:kolyadin-spectral-loss-profile}.

The studies of this section have shown that inclusion of a lossy
polymer layer smooths out the variations in loss profile. 
These observations suggest that carefully modeling polymer layers will likely be significant when experimentally validating 
numerically predicted mode loss profiles. 
Conversely, not precisely knowing the material properties of the polymer coatings, may also be a limitation to experimental validation. 
Next, we will see that potential, but expected, sensitivities to variation in fiber geometry, may or may not appear in the CL loss profiles, depending on the modeling choices regarding the fiber's outer cladding layers.

\section{Sensitivity to geometric parameters in $C_0$ and $C_1$ models}
\label{sec:emb}

It is natural to investigate how susceptible the computed losses of a microstructured fiber are to expected geometrical variations of the design. 
Ideally, one desires robust designs, tolerant to typical fabrication-induced variances. 
In \cite[Figure~9]{polettiNestedAntiresonantNodeless2014}, the authors
considered the effect of varying the azimuthal separation of the capillary tubes on the spectral loss profiles of their novel fiber design.
These hollow tubes were allowed to vary in diameter while keeping the core
diameter fixed, resulting in both changes in azimuthal separation of the tubes and in
the location of the glass cladding that supports these capillaries. 
They showed that varying this parameter changes both the minimum achievable loss value and the bandwidth of the loss profile, i.e., the range of wavelengths for which the loss remains low. 
The loss profiles changed dramatically for small initial changes to the azimuthal
distance, from $0$ to $0.84  \micrometer$.  
  
\begin{figure}[b]
     \centering
	\begin{tikzpicture}
\colorlet{mygreen}{green!70!blue}
    \begin{groupplot}[
        group style={
            group size=1 by 2,
            vertical sep=.8cm,
            xlabels at=edge bottom,
        },
	embedding,
    xlabel={Embedding Fraction ($e_{cap} / t_{cap}$)},
    ylabel={Loss (dB/m)},
    ymode=log,
    log ticks with fixed point,
    enlarge x limits=.02,
    enlarge y limits=.05,
    height=.42\textwidth,
    legend style={legend pos=north west, font=\tiny},
	]
    \nextgroupplot[xtick={0.0, .1,.2,.3,.4,.5,.6,.7,.8,.9,1.},
    width=.95\textwidth,
    height=.4\textwidth,
     ]
	\addplot+[
	    color=BrickRed,
	    mark=+,
	    mark size=0pt,
	    line width=.85pt,
	    ]
	table[]{figures/data/embed/k_inf.dat};
	\addlegendentry{$C_0$}
	\addplot+[
	    color=blue,
	    mark=+,
	    mark size=0pt,
	    line width=.85pt,
	    ]
	table[]{figures/data/embed/k_0.0.dat};
	\addlegendentry{$C_1$}
    \nextgroupplot[xtick={0.0, .1,.2,.3,.4,.5,.6,.7,.8,.9,1.},
        width=.95\textwidth,
    height=.4\textwidth,
     ]
	\addplot+[
		color=Blue!15,
	    mark=+,
	    mark size=0pt,
	    line width=1.1pt,
	    ]
	table[]{figures/data/embed/k_0.005.dat};
	\addlegendentry{$k=.005$}
	\addplot+[
	    color=Blue!25,
	    mark=+,
	    mark size=0pt,
	    line width=1pt,
	    ]
	table[]{figures/data/embed/k_0.001.dat};
\addlegendentry{$k=.001$}
	\addplot+[
	    color=Blue!40,
	    mark=+,
	    mark size=0pt,
	    line width=.9pt,
	    ]
	table[]{figures/data/embed/k_0.0005.dat};
	\addlegendentry{$k=.0005$}
	\addplot+[
	    color=Blue!70,
	    mark=+,
	    mark size=0pt,
	    line width=.8pt,
	    ]
	table[]{figures/data/embed/k_0.0001.dat};
	\addlegendentry{$k=.0001$}
	\addplot+[
	    color=Blue,
	    mark=+,
	    mark size=0pt,
	    line width=.7pt,
	    ]
	table[]{figures/data/embed/k_1e-05.dat};
	\addlegendentry{$k=.00001$}
\end{groupplot}
\end{tikzpicture}
 \caption{Embedding Sensitivity, 6-tube ARF.}
\label{fig: embed plots}
\end{figure}
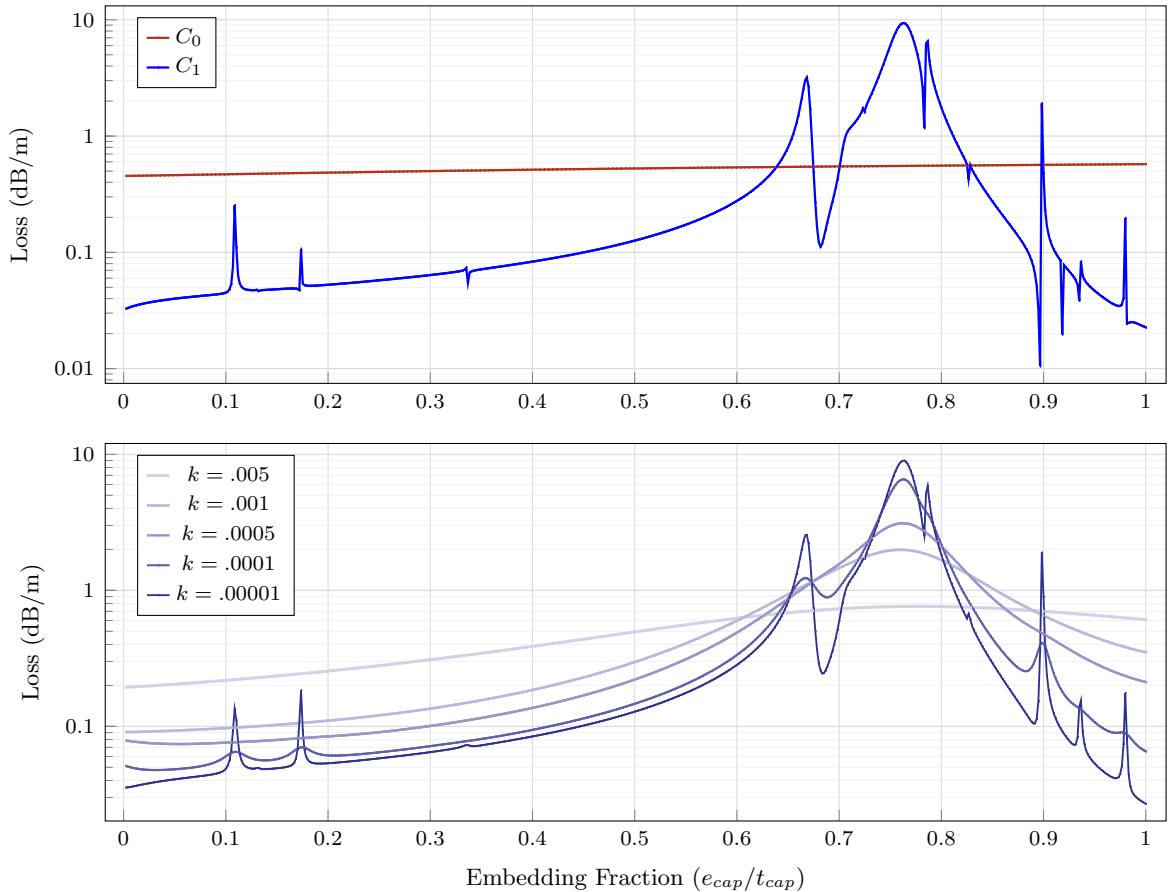

In a similar fashion, we study the effects of varying the radial embedding distance of capillary tubes into their supporting glass cladding; an expected geometrical variation to the fiber design. 
Realistically, the microstructure cannot meet the outer cladding in a perfectly tangential manner (zero embedding distance).
Rather, some amount of capillary tube material must be absorbed into the support. 
This is modeled by allowing the tubes to embed some distance $e_{cap}$ (see Figure~\ref{fig: arf geom}) into the cladding, varying the distance between 0 and the thickness of the capillary tube. 
We keep the core region and capillary tube radii constant, which implies that the radius to the start of the glass cladding shifts to achieve the desired embedding.

The upper plot of Fig.~\ref{fig: embed plots} shows how CL values change as the embedding fraction ($ e_{cap} / t_{cap}$) is varied from 0 to 1 for the 6-tube ARF design. 
The results strongly depend on the choice of outer material layer configurations. 
In a $C_0$ model, where the glass cladding extends to infinity, the CL values exhibit essentially no sensitivity to changes in embedding fraction. 
Whereas for a $C_1$ configuration, where the ambient material is air, one observes that the CL values vary by about three orders of magnitude, which would have drastic impacts on optical field propagation in such a waveguide. 
The lower plot of Fig.~\ref{fig: embed plots} illustrates the numerical results for the case of a lossy polymer layer (for various extinction coefficients~$k$) between the glass cladding and the infinite layer of air. 
Again, one observes loss profiles that act as a gradient between the two limiting cases as was demonstrated for the polymer coated Bragg fiber design, in subsection~\ref{subsec:bragg-poly}.

These results suggest that if the fiber coating has a small enough extinction coefficient
at the operating optical wavelength, then the embedding distance of the capillary tubes may
play a large role in determining the mode loss. 
For higher polymer coating loses, perturbations in geometry will not affect the computed mode loss. 
However, if the loss in the polymer constitutes an absorption (rather than scattering), then one may find sufficient levels of thermal heating to induce polymer degradation/destruction~\cite{snyderOpticalWaveguideTheory2024}. 
Overall, this is a demonstration that emphasizes the fact that the outer material layer configurations must be treated correctly in the model to assess the robustness of a fiber design to the expected tolerances in geometry. 
Potentially, this outer layer treatment issue may explain previously unknown sources of loss and/or polymer heating in hollow-core/anti-resonant fiber designs.

\section{Accelerating convergence with finer mesh in high index regions}
\label{sec:accel-conv}

It has been reported~\cite{gopalakrishnanComputingLeakyModes2022, polettiNestedAntiresonantNodeless2014}
that numerical modeling of ARF fibers requires extremely
fine mesh sizes for accurate loss measurements.  In
\cite{polettiNestedAntiresonantNodeless2014}, the use of quadratic finite
elements with element sizes at most $\lambda/4$ in the air and $\lambda/6$ in
the glass regions was found to be necessary.  Here we report that convergence
of CL values for these fibers may be accelerated by reducing the element size
in the glass portions of the fiber alone. While there is broad agreement that meshsizes must be small enough to capture the thin curvilinear
capillary structures, what may be non-intuitive is that the glass cladding region without any tiny geometrical features also needs a fine mesh, as we shall reason
below.
\begin{figure}[b]
     \centering
     \begin{subfigure}[b]{.35\textwidth}
         \centering
          \adjincludegraphics[width=.95\textwidth, height=.65\textwidth, trim={6.5in 5in 7in 4.75in}, clip]{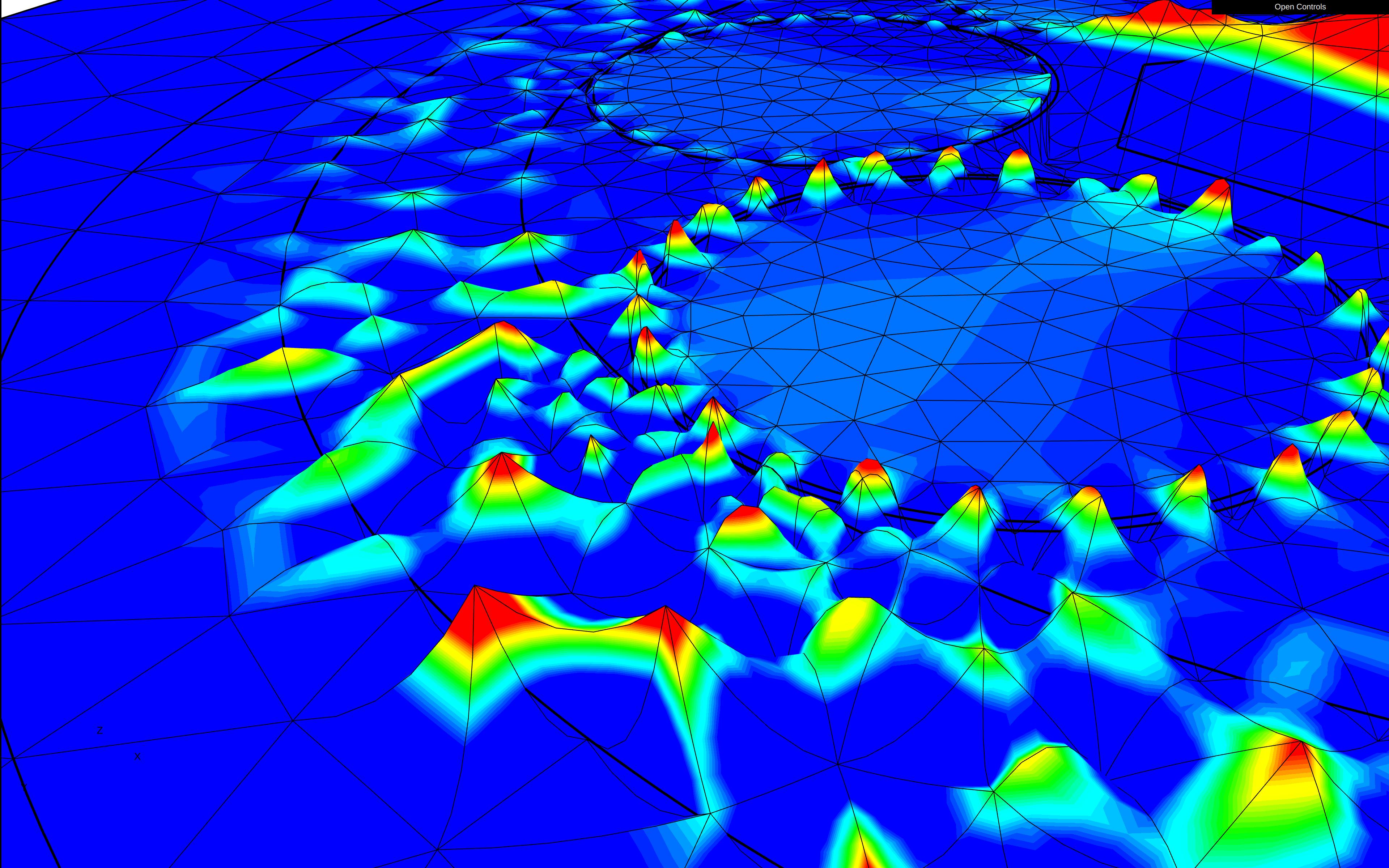}
          \caption{$\lambda/h \approx 0.3$}
     \end{subfigure}
     \hspace{.2in}
     \begin{subfigure}[b]{0.35\textwidth}
         \centering
         \adjincludegraphics[width=.95\textwidth, height=.65\textwidth, trim={7in 5in 9in 5in}, clip]{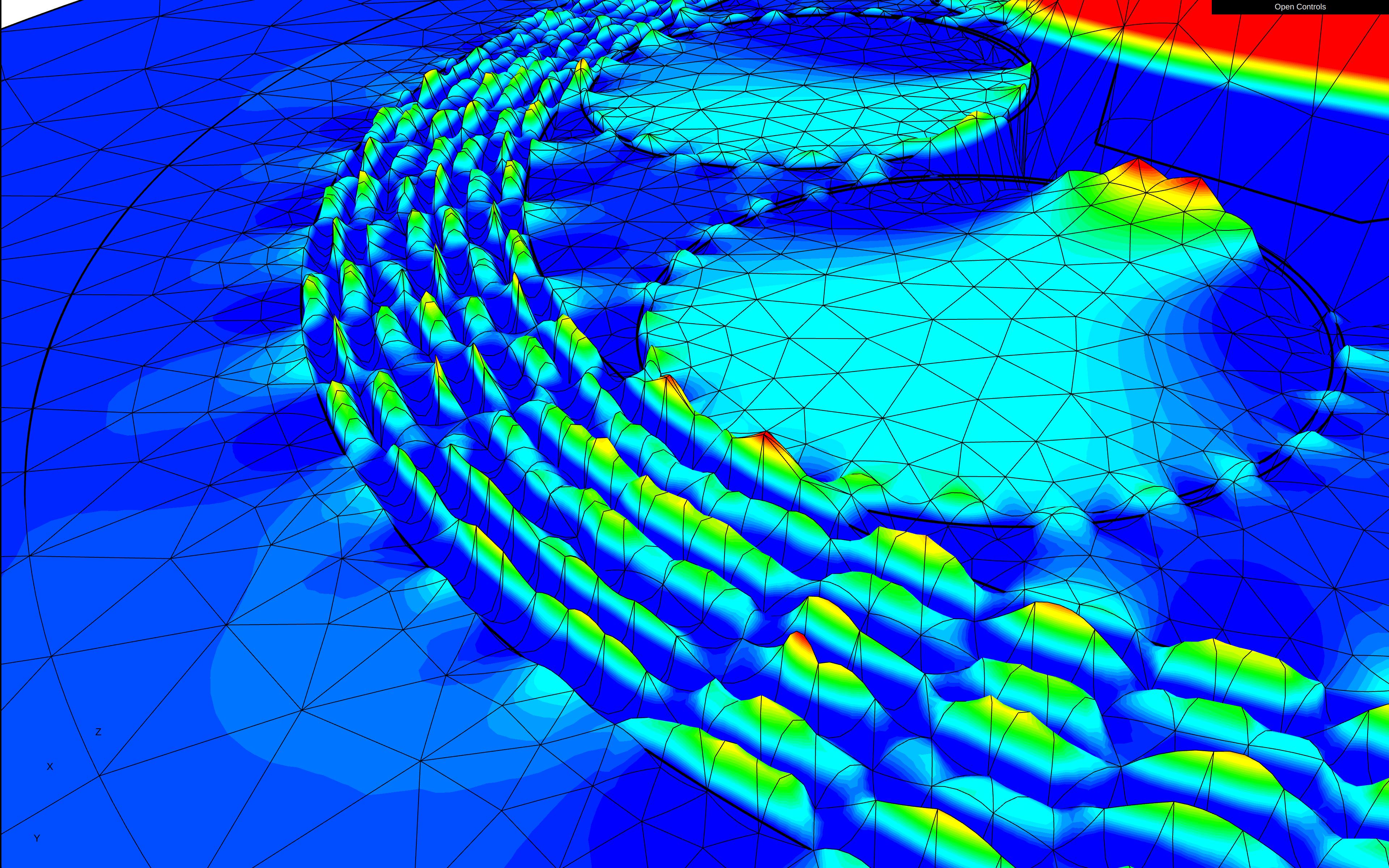}
         \caption{$\lambda/h \approx 0.9$}
      \end{subfigure}\vspace{.1in}
      \begin{subfigure}[b]{.35\textwidth}
         \centering
          \adjincludegraphics[width=.95\textwidth, height=.65\textwidth, trim={7in 5in 9in 5in}, clip]{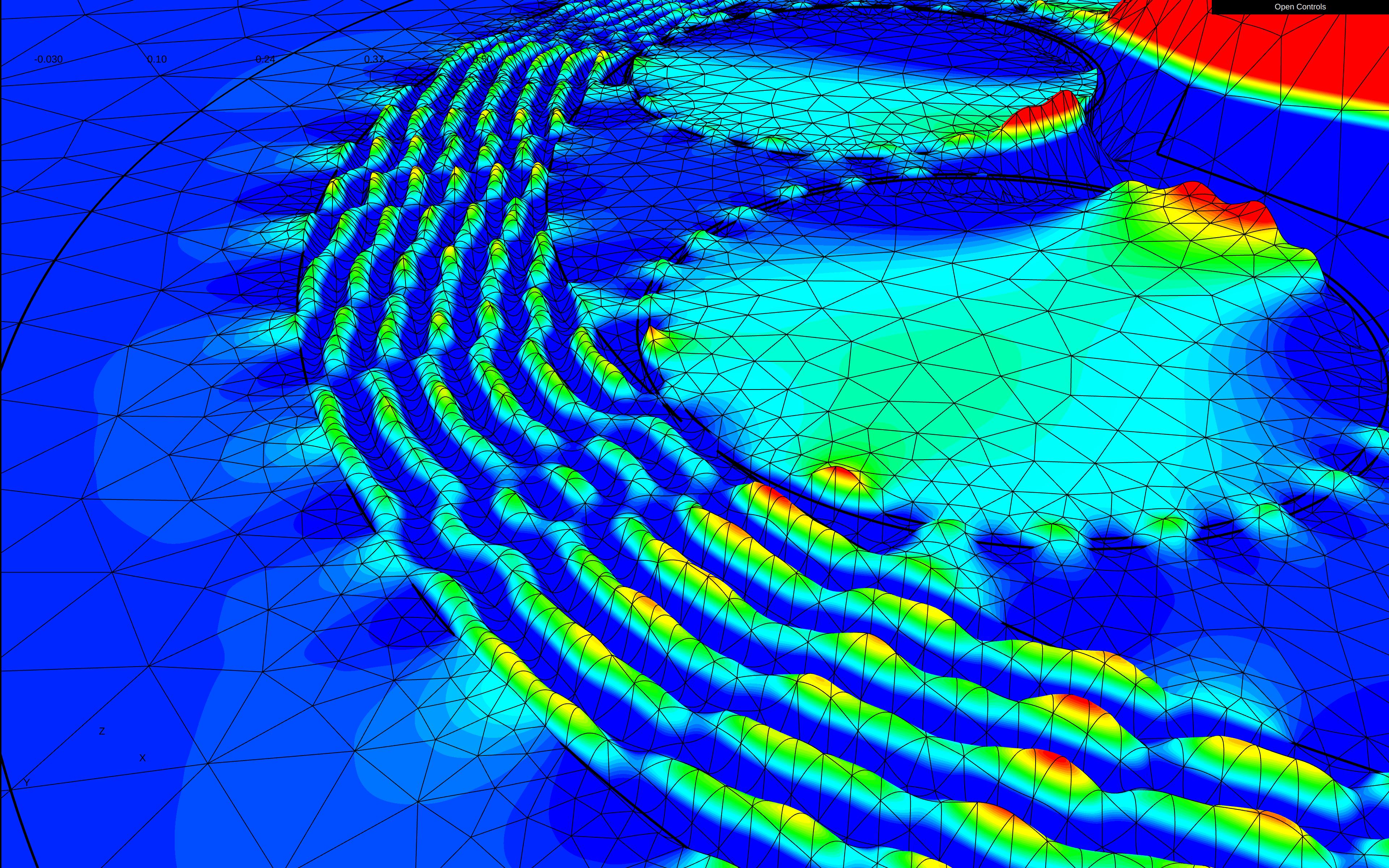}
         \caption{$\lambda/h \approx 1.5$}
     \end{subfigure}
     \hspace{.2in}
     \begin{subfigure}[b]{0.35\textwidth}
         \centering
         \adjincludegraphics[width=.95\textwidth, height=.65\textwidth, trim={7in 5in 9in 5in}, clip]{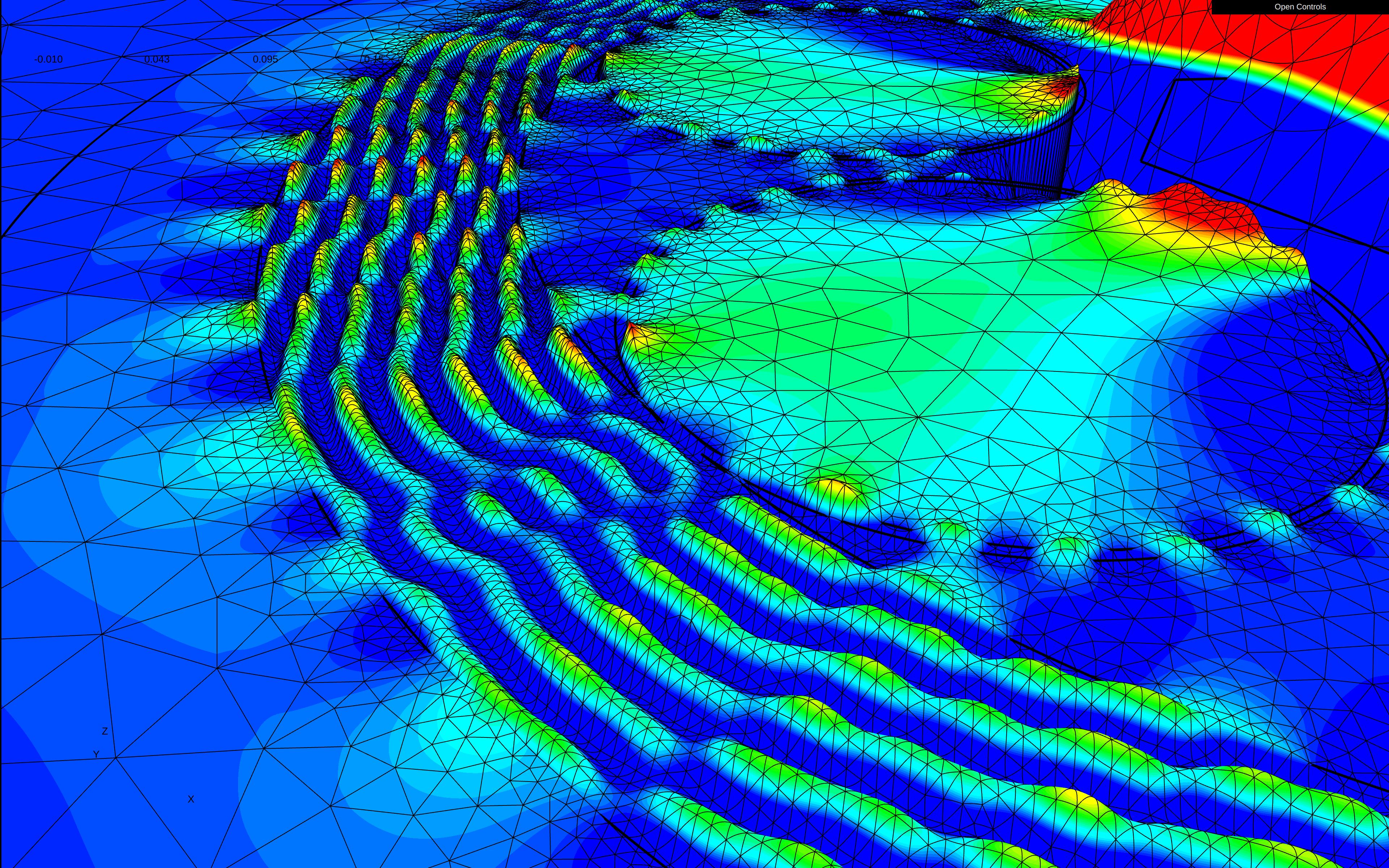}
         \caption{$\lambda/h \approx 3.0$}
      \end{subfigure}
 \caption{Longitudinal component of fundamental mode for the 6-tube fiber 
 in cladding region, found with
 order 3 elements, on four meshes. }
\label{fig: mesh mode compare}
\end{figure}

Through a careful examination of the computed mode profiles, we
found that in the cladding region, a fine radial wave is present, even
at wavelengths where the fiber is in anti-resonance.  This is in line
with the mode profile ripples (standing waves) seen in the high index 
regions of the Bragg fiber designs.  
When using a coarse mesh in this region, it is not possible
to capture this fine wave pattern accurately with low order elements.
Decreasing the meshsize in this region allows these ripples to be
better approximated with elements of the same order.  The amplitude of
these ripples is often much smaller in magnitude than that of the field 
value in the core region, and hence
it may be easily missed in a quick first look at the mode.  To clearly
portray this phenomenon, we include a zoomed-in view of a mode profile,
near a capillary structure, computed on a series of four meshes from
coarse to fine; see Figure~\ref{fig: mesh mode compare}.  To obtain the
finer meshes, we decreased the meshsize only in the cladding region,
with element size changing in low index regions only as necessary to
maintain geometrical conformity of elements at the
interfaces. Clearly, the ripples are already well approximated in the
third mesh and the fourth mesh confirms that ripples (are not a
numerical artifact and) remain unchanged under further refinements.
We hypothesize that capturing this fine ripple structure in the
cladding is critical for quick numerical convergence of CL values.

\begin{figure}
\begin{subfigure}[b]{\textwidth}
\centering
\pgfplotstableread{figures/data/mesh_maxh/coarse_cladding1.dat}\coarseA
\pgfplotstableread{figures/data/mesh_maxh/coarse_cladding2.dat}\coarseB
\pgfplotstableread{figures/data/mesh_maxh/maxh_0.1_ref0.dat}\fineA
\pgfplotstableread{figures/data/mesh_maxh/maxh_0.1_ref1.dat}\fineB

\SetElement{\coarseA}{0}{1}{\ymin}
\SetElement{\coarseA}{0}{0}{\xmin}
\SetElement{\fineA}{0}{1}{\ymax}
\SetElement{\coarseA}{10}{0}{\coarseL}
\SetElement{\fineA}{5}{0}{\fineL}

\colorlet{myblue}{Blue!100}
\colorlet{myorange}{orange!90!blue}

\begin{tikzpicture}
    \begin{axis}[
        convergence_narrow,
        width=.9\textwidth,
        height=0.4\textwidth,
	    xlabel={Degrees of freedom},
	    ylabel={Loss (dB/m)},
	    ymode=log,
	    xmode=log,
		ymin=3*10^-7,
		ymax=9,
		enlarge y limits=0,
		enlarge x limits=.04,
		legend style={legend pos=south east, font=\tiny},
    	title style={at={(0,1)},anchor=north west, yshift=-6},
    	title={},			 
		]
		\addplot +[mark=none, color=myorange, dashed, line width=1.2pt,
						  forget plot] coordinates {(\coarseL, 10^-8) (\coarseL, 100)};
		\addplot+[
		    color=myorange,
		    mark=*,
		    mark options={fill=orange!80!blue,},
		    mark size=1pt,
		    line width=.85pt,
		    ]
		table[]{figures/data/mesh_maxh/coarse_cladding1.dat};

			\addlegendentry{\tiny Naive}
			\addplot+[
			    color=myblue,
			    mark=*,
			    mark options={fill=myblue},
			    mark size=.8pt,
			    line width=1pt,
			    ]
			table[]{figures/data/mesh_maxh/maxh_0.1_ref0.dat};
			\addplot +[mark=none, color=myblue, dashed, line width=1.2pt,
							 forget plot] coordinates {(\fineL, 10^-8) (\fineL, 100)};
			\addlegendentry{\tiny Informed}
\end{axis}
\end{tikzpicture}
\caption{Convergence for naive and informed meshes}
\label{fig:naive-inform-conv-compare}
\end{subfigure}\vspace{.15in}
     \hspace*{.6in}\begin{subfigure}[b]{.42\textwidth}
         \centering
          \adjincludegraphics[height=.65\textwidth, trim={12.5in 5in 11in 4.75in}, clip]{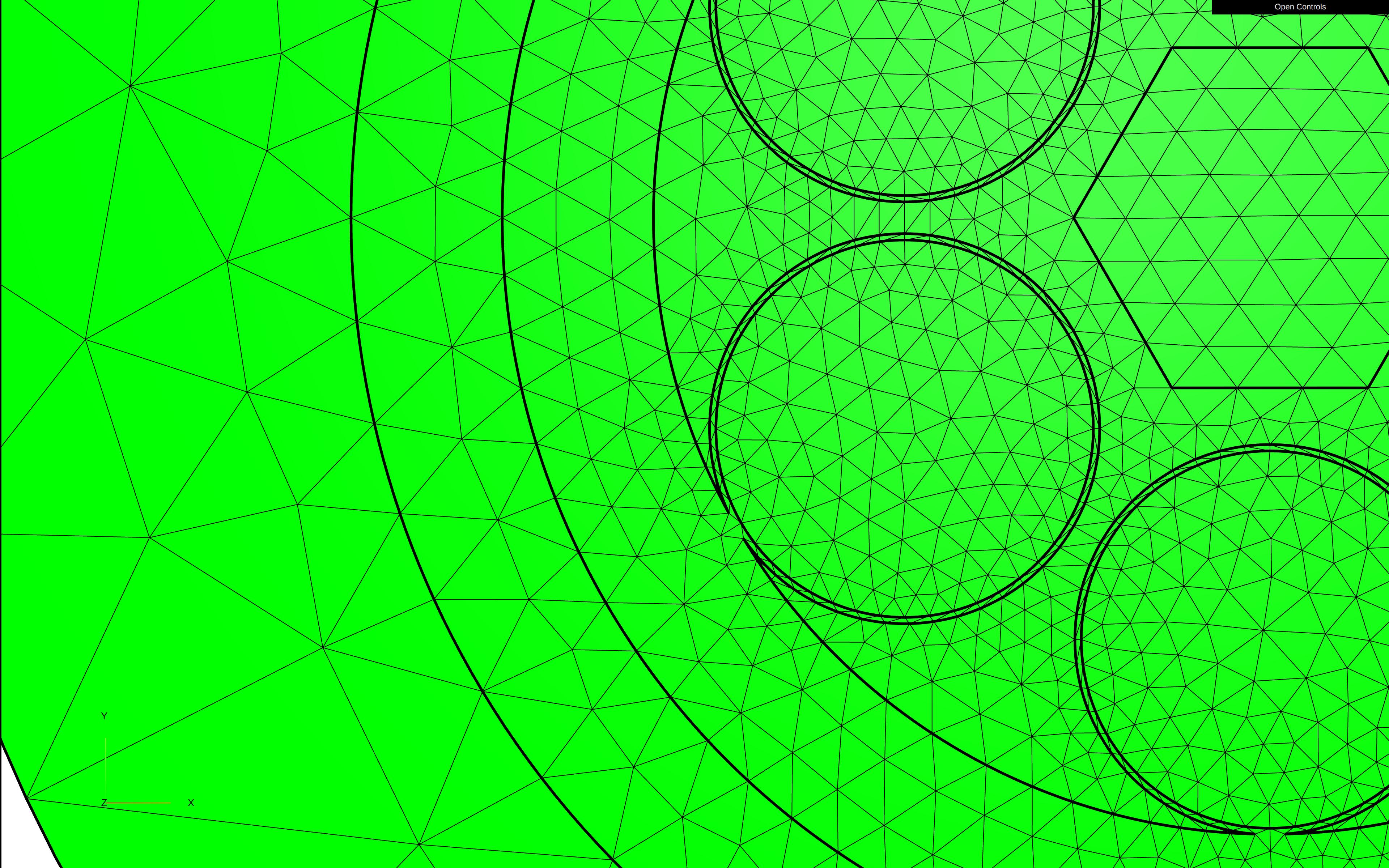}
          \caption{Naive mesh: $\lambda/h \approx .3$ in cladding}
          \label{fig:cgce-naive}
     \end{subfigure}
     \hspace{.3in}
     \begin{subfigure}[b]{0.42\textwidth}
         \centering
         \adjincludegraphics[height=.65\textwidth, trim={15in 6in 8.5in 3.75in}, clip]{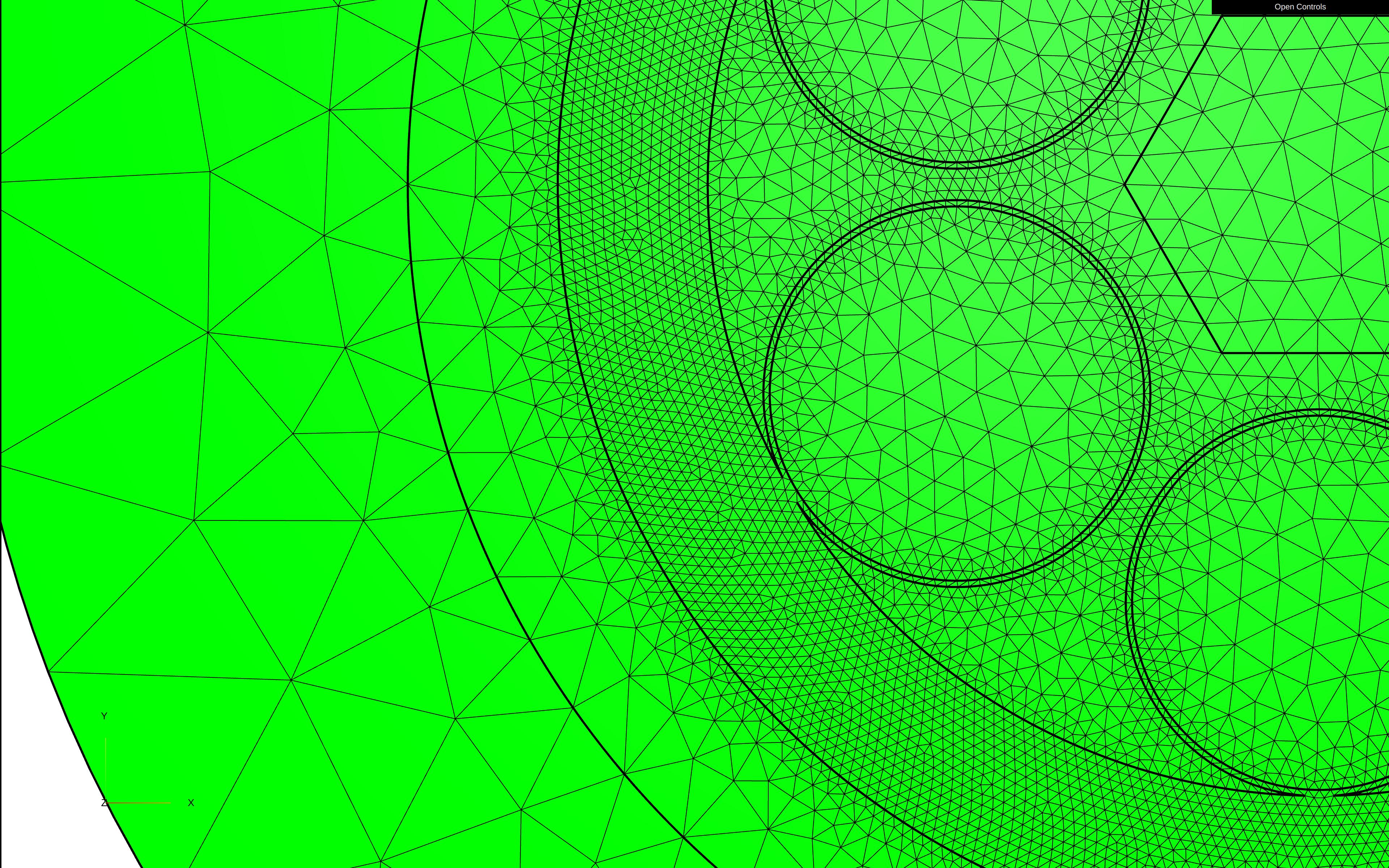}
         \caption{Informed mesh: $\lambda/h \approx .9$ in cladding}
         \label{fig:cgce-inform}
       \end{subfigure}
       \caption{Comparison of convergence for naive and informed meshes.}
\label{mesh}
\end{figure}
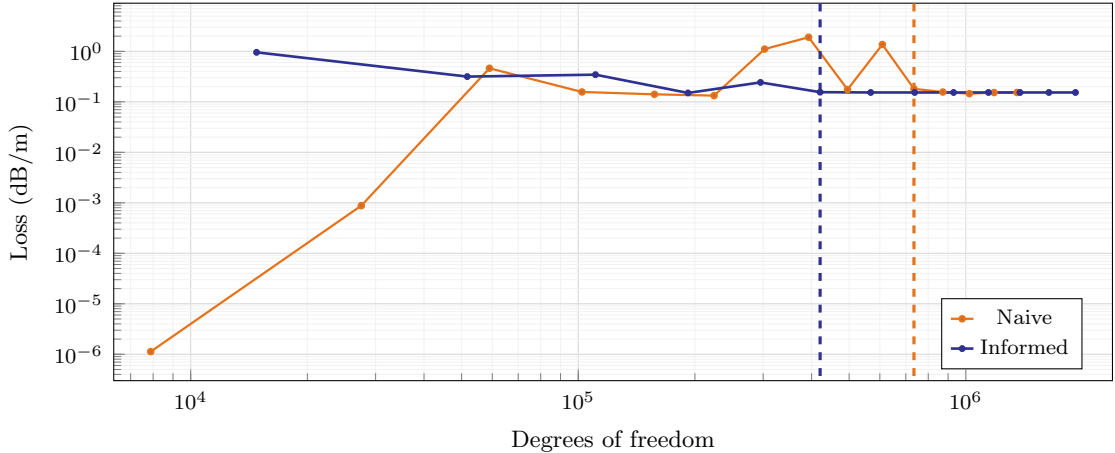
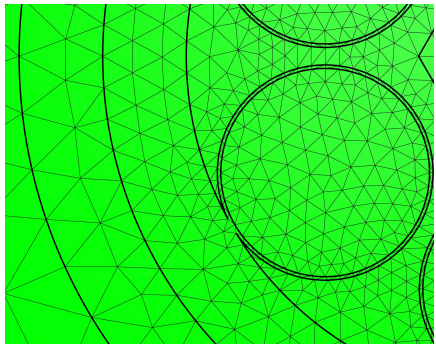
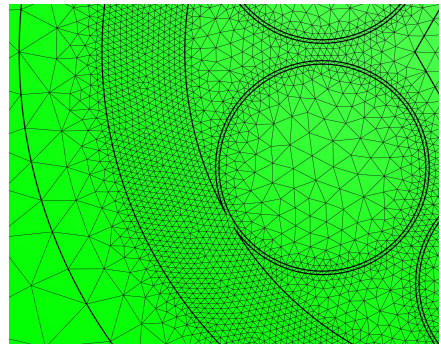

To test this hypothesis, we conducted computations starting from two
meshes, a ``naive'' mesh with large elements in the cladding, and
an ``informed'' mesh with fine elements in the cladding, both shown in
Figure~\ref{mesh}. Mesh sizes in
all other regions are kept relatively coarse, except for changes needed at the
interfaces between high and low index materials.
The maximum diameter of elements in low index regions was kept near $\lambda/0.4$
elements per wavelength.  
We computed CL values for a sequence of increasing
element orders, from linear, to quadratic, and higher. Specifically,
the corresponding degree
$p$, as described in \S\S\ref{sec:discret}, 
goes from $p=0 \ldots 14$ on the mesh shown in Figure~\ref{fig:cgce-naive} and 
from $p=0 \ldots 12$ on the mesh shown in Figure~\ref{fig:cgce-inform}.  Vertical 
lines indicate the locations where each convergence plot enters
the regime of converged CL values.

For all of the meshes, we see numerical convergence
as we increase the order of the elements used. However, we see that for  the informed mesh with 
finer elements in the high index regions, the CL values converge quicker and there is little need to go beyond order six or seven. For the naive mesh, variations in computed CL values persist even as we near a million degrees of freedom.
In other words, using a naive mesh, one faces a large preasymptotic regime, which must be crossed before trustworthy convergent CL values can be computed.
The presence of such a large preasymptotic regime for ARF fiber models and the importance of crossing it was first reported in~\cite{gopalakrishnanComputingLeakyModes2022}. There, 
in \cite[Fig.~8]{gopalakrishnanComputingLeakyModes2022}, 
a plot similar to Figure~\ref{fig:cgce-naive} can be
found using results from a scalar mode solver (while Figure~\ref{mesh} is made using results from the hybrid vector mode solver in Appendix~\ref{sec:numerical-method}).
In this section, by identifying a potential reason for the large
preasymptotic regime, we have shown one technique to make it significantly smaller by using the informed mesh, allowing us to get convergent CL values with fewer degrees of freedom. 

\section{Discussion}
\label{sec:discussion}

In this work it has been established that confinement loss values for leaky modes of 
anti-resonant fibers depend strongly on the choice of material configuration at the
computational boundary.  Allowing the cladding material to extend to the boundary
leads to smooth loss profiles, while terminating the cladding and immersing the 
geometry in low index material leads to highly varying loss profiles.  Inclusion of
a lossy polymer layer led to profiles between these two extremes, with the variability
of the profile decreasing as material loss increased.

Because of this strong dependence, we recommend that numerical studies involving optical 
fiber modes provide details of any included outer material layers
and of the implementation of the computational boundary, as it is now clear that this 
information is essential for reproducibility. This information further allows 
for others to make comparisons with different modeling choices and/or later
experimental measurements.

The establishment of this dependence also raises the possibility that modelers
may be able to more accurately predict experimental results by including complete
representations of coating materials into their optical fiber models.  It is our
hope that this work inspires experiments exploring this possibility.  Such experiments
would require detailed information on the wavelength dependent material 
losses of included coating materials. Ideally, polymer
coating manufacturers would release material loss values over relevant
optical wavelength ranges for their materials. More public domain
information on these materials could also lead to further studies on energy
conversions (such as optical to thermal energy) within polymers and
how that in turn affect mode losses.



When initially designing an optical fiber, information on coatings may not be present.
If a fiber is at this stage of development, we recommend that modelers 
investigate losses using both configurations.   The loss profile found using 
the $C_0$ model contains core information about the fiber design.  This is attested to by the fact that 
it is the limiting profile in the polymer studies of section \ref{sec:polymer}.   
The loss profile found with the $C_1$ model expands on this information, showing what wavelength
regions may later have high variability experimentally.  Modelers may be able to use this
information to change the fiber design in ways that smooth the final experimental loss profile.

%

%

We conclude by offering a few general recommendations gathered from
our numerical experience with modern antiresonant fibers.  First, one
must recognize the importance of not presuming a continuous, nor
differentiable, refractive index profile, particularly for
microstructured fibers (which are not weakly guiding) with large index
contrast.  The importance of choosing the correct mode solving approach is
emphasized at the beginning of Appendix~\ref{sec:numerical-method}.
Furthermore, now that it has been established that the mode profiles do exhibit standing waves (ripples) in high-index cladding regions (anti-resonant layers), 
we encourage modelers to follow the informed meshing guidelines that we presented in Section~\ref{sec:accel-conv}.
Finally, we recommend that researchers ensure  
numerical convergence for their computed mode data, as this may require higher degrees of freedom than one might initially expect.  

\appendix

\section{The eigenproblem and the numerical method}
\label{sec:numerical-method}
Simulation techniques for hybrid vector modes of optical fibers are
based on formulations that fall into two categories. The first uses
derivatives of the refractive index in the equations: examples
include~\cite{snyderModesOpticalWaveguides1978, monroModelingLargeAir2000,
  uranusModellingMicrostructuredWaveguides2004}. The method we 
  describe below falls into the second
category of formulations without index
derivatives~\cite{koshibaSimpleEfficientFiniteelement1992,
koshibaCurvilinearHybridEdge2000, leeFullwaveAnalysisDielectric1991, 
vardapetyanFullWaveAnalysisDielectric2003}.
Note that the first category is unsuitable for application to modern
microstructured fibers with jumps in refractive index: such index
functions are not differentiable (not even continuous).

\subsection{Constrained eigenproblem for hybrid vector modes}
\label{ssec:constr-eigen}

Consider the time-harmonic Maxwell system for the electric field
$\hat E$ and the magnetic field $\hat H$, assuming that time
variations are of the form $e^{-\ii \og t}$, namely 
\begin{align}
  \label{eq:1}
  -\ii \og \mu \hat H + \nabla \times \hat E
  & = 0, 
  & \ii \og \veps \hat E + \nabla \times \hat H & =0, 
\end{align}
for all $(x, y, z) \in \R^3$. The $xy$-plane represents the transverse
cross section of an optical fiber placed along $z$-axis.  The fiber
may have complex transverse microstructure, but has translational
symmetry in the $z$-direction.  In~\eqref{eq:1}, $\nabla \times \cdot$
denotes the three-dimensional curl, while 
$\veps$ represents 
the electric permittivity 
and $\mu$ represents the 
magnetic permeability.  We assume that $\mu>0$
is isotropic and constant and that $\veps$ takes the form
\begin{equation}
  \label{eq:eps-block}
  \veps =
  \begin{bmatrix}
    \veps_{xx}(x, y) & \veps_{xy}(x, y) & 0 \\
    \veps_{xy}(x, y) & \veps_{yy}(x, y) & 0 \\
    0 & 0 & \veps_z(x, y)
  \end{bmatrix}
  \equiv
  \begin{bmatrix}
    \veps_\tau & 0 \\
    0 & \veps_z
  \end{bmatrix},
  \quad \text{ with }
  \veps_\tau =
  \begin{bmatrix}
    \veps_{xx} & \veps_{xy} \\
    \veps_{xy} & \veps_{yy} 
  \end{bmatrix},  
\end{equation}
with
symmetric positive definite $\veps_\tau$
and  $\veps_z> 0$  at every point
$(x, y)$ of the transverse plane. Nonconstant, even discontinuous, 
and anisotropic~$\veps$ are included under these assumptions.

We are interested in Maxwell solutions that propagate along the
fiber's longitudinal direction. Accordingly, we seek solutions of the
form $ \hat E(x, y, z) = E(x, y) e^{\ii \beta z}$ and
$ \hat H(x, y, z) = H(x, y) e^{\ii \beta z}, $ where 
$\beta$ is the propagation constant. The fields   $E, H$ are {\em hybrid
  modes}, i.e., vector fields with both transverse and longitudinal
components, all of which depend only on the transverse coordinates $x, y$.
Using the unit coordinate vectors, $e_x, e_y,$ and $e_z$, in the $x,y$
and $z$ directions, respectively, we may decompose a hybrid field $u$
into $u(x,y) = u_x(x,y) e_x + u_y(x,y) e_y + u_z(x,y) e_z.$ Denote its
transverse projection by $ u_\tau = u_x e_x + u_y e_y$.  Let
$J = \left[\begin{smallmatrix} \phantom{-}0 & 1 \\ -1 & 0
\end{smallmatrix}\right]
$ denote the operator that rotates
vectors in the $xy$-plane clockwise by 90 degrees about the positive
$z$-axis. For transverse fields, we define the 2D (scalar)
curl and (vector) rot by
$
\curl u_\tau = \dive (J u_\tau) = \d_x u_y - \d_y u_x,$ and $
\rot \phi = J (\grad \phi) = (\d_y\phi) e_x - (\d_x \phi) e_y.  
$
Using the identity
$  \nabla \times \left(u\, e^{\ii\beta z}\right) =
\left( e^{\ii\beta z} \curl u_\tau\right)e_z + 
e^{\ii\beta z}\left( \rot u_z - \ii\beta J u_\tau \right), 
$
the system~\eqref{eq:1} simplifies
into the following  coupled system for the two transverse vector fields
$E_\tau, H_\tau,$ the two longitudinal components $E_z, H_z,$ and the
associated propagation constant $\beta$. 
\begin{subequations}
  \label{eq:EHfullsystem}
  \begin{align}
    \label{eq:EHfullsystem-1}
    -\ii\og \mu H_\tau + \rot E_z & = \ii \beta J E_\tau,
    & 
    -\ii\og \mu H_z + \curl E_\tau & = 0,
    \\  \label{eq:EHfullsystem-3}
    \ii\og \veps_\tau E_\tau + \rot H_z & = \ii\beta J H_\tau, 
    & 
    \ii\og\veps_z E_z + \curl H_\tau & = 0.
  \end{align}
\end{subequations}

Next, we eliminate the variables $H_z$ and $H_\tau$
from~\eqref{eq:EHfullsystem} and simplify to obtain the following
system for $E_\tau$ and the scaled longitudinal component
$\vphi = \ii \beta E_z$,
\begin{subequations}
  \label{eq:5}
  \begin{align}
    \label{eq:5a}
    \rot\curl E_\tau - \og^2 \veps_\tau \mu E_\tau + \grad \varphi
    & = - \beta^2 E_\tau,
    \\
    \label{eq:5b}
    -\Delta_\tau \varphi - \og^2 \veps_z \mu \varphi & = \beta^2 \dive E_\tau.
  \end{align}
\end{subequations}
Let $\veps_0$ and $\mu_0$ denote the electric permittivity and
magnetic permeability of vacuum and let $k^2 = \og^2 \veps_0 \mu_0$.
Define the transverse refractive index $n_\tau(x, y)$ as the unique
$2 \times 2$ symmetric positive definite matrix satisfying
$n_\tau^2 = \veps_\tau \mu/(\veps_0\mu_0)$,
let $n_z = \sqrt{\veps_z \mu /(\veps_0\mu_0)}$, and let
$n(x, y) =
\left[
\begin{smallmatrix}
    n_\tau & 0 \\
    0 & n_z
\end{smallmatrix}
\right]$ in the same block partitioning as in \eqref{eq:eps-block}.  Taking
the transverse divergence of both sides of~\eqref{eq:5a} and adding it
to~\eqref{eq:5b}, we see that
$k^2 n_z^2 \varphi + \dive (k^2 n_\tau^2 \varphi) = 0
$. Thus~\eqref{eq:5} is equivalent to the following {\em eigensystem
  with a constraint} for the eigenvalue $-\beta^2$:
\begin{subequations}
  \label{eq:6}
  \begin{align}
    \label{eq:6a}
    \rot\curl E_\tau - k^2n_\tau^2 E_\tau + \grad \varphi & = - \beta^2 E_\tau,
    \\
    \label{eq:6b}
    n_z^2 \varphi + \dive (n_\tau^2 E_\tau) & = 0.
  \end{align}
\end{subequations}

\subsection{Nondimensionalization}
\label{ssec:nondim}

Considering an optical field in the infrared wavelength regime and 
practical fiber geometries, one finds propagation constants $\beta$ of the
order of millions, with a length scale in micrometers.  Hence
nondimensionalization is necessary for preserving digits of accuracy
in computations.  Outside some finite radius in the transverse plane,
the refractive index $n(x, y)$ is isotropic and equal to a constant $n_0$.
Define the {\em index well}\;
$ V(x, y) =  k^2 (n_0^2  - n_\tau(x, y)^2).
$
Next, using a length scale $L$ appropriate for the fiber cross section, 
we introduce the
nondimensional transverse coordinates $ \tilde x = x/L, $ and
$ \tilde y = y/L, $ and 
set 
\[
  \begin{gathered}
  \tilde E_\tau (\tilde x, \tilde y) = E_\tau  (L\tilde x, L\tilde y),
  \quad
  \tilde \varphi (\tilde x, \tilde y) = L \varphi(L\tilde x, L\tilde
  y),
  \\
  \tilde n_\tau (\tilde x, \tilde y) = n_\tau  (L\tilde x, L\tilde y),
  \quad
  \tilde n_z (\tilde x, \tilde y) = n_z  (L\tilde x, L\tilde y),
  \quad
  \tilde V (\tilde x, \tilde y) = L^2 V(L\tilde x, L\tilde y).
  \end{gathered}
\]
Then~\eqref{eq:6}
implies the following nondimensional system
\begin{subequations}
  \label{eq:7}
  \begin{align}
    \label{eq:7a}
    \rot \,\curl \tilde{E}_\tau +  \Vt \tilde{E}_\tau +
    \grad \tilde\varphi & = Z^2 \tilde{E}_\tau,
    \\
    \label{eq:7b}
    \tilde{n}_z^2 \tilde\varphi + \dive (\tilde{n}_\tau^2 \tilde{E}_\tau) & = 0,
  \end{align}
\end{subequations}
where the derivatives are now computed with respect to the nondimensional 
$\tilde x$ and $\tilde y$, and
$  Z^2 = L^2 (k^2 n_0^2 - \beta^2)$
takes the role of the corresponding nondimensional eigenvalue.  
We henceforth drop the
tildes from the quantities in~\eqref{eq:7} since we always compute with the  nondimensionalized system.

\subsection{Discretization using \Nedelec and Lagrange finite elements}
\label{sec:discret}

For guided modes, we use the equations of~\eqref{eq:7} on a finite
domain (a large enough circle) $\om$ in the $xy$-plane, together with the boundary condition
$ E_\tau = \vphi = 0$ on $\d\om$. This condition is
appropriate when $\d\om$ is far enough away from the compact support of $V$
for the guided mode components to have exponentially decayed to machine zero.  Let
$\om$ be meshed by a geometrically conforming finite element mesh of
triangles, denoted by $\oh$. On a triangle $K$, let $P_p(K)$ denote
the space of polynomials in two variables of degree at most~$p$. The
degree $p$ and the maximal element diameter
$h = \max_{K \in \oh} \mathrm{diam}(K)$ together can be used to
compute the dimension of the discrete spaces.  The Lagrange and the
\Nedelec\ spaces on $\oh$ are defined, respectively, for any $p\ge 0$
by
$\Vhp = \{ \psi: \psi|_K \in P_{p+1}(K) \text{ for all } K \in \oh,
\text{ $\psi$ is continuous} \},$ and
$ \Nhp = \{ v \in \Ho(\curl,\om): v|_K \in P_p(K)^2 +
\left[\begin{smallmatrix} \phantom{-}y \\ -x
  \end{smallmatrix}\right]
P_p(K) \text{ for all } K \in \oh, \text{$v$ is tangentially
  continuous}\}.$ Adjacent to curved material interfaces, we use
curved triangles, in which case, the space within such an element is
revised as usual to equal the pullback of the above-indicated
polynomial spaces from a unit reference triangle.

Using both the boundary conditions $E_\tau =0$ and $ \vphi = 0$ after multiplying~\eqref{eq:7} by test
functions and integrating by parts, we obtain a ``mixed finite
element discretization'' \cite{monkFiniteElementMethods2003}
 that computes approximations
of $E_\tau$ and $\vphi$, denoted by $\Eth \in \Nhp$ and $\vph\in \Vhp$,
respectively, by solving 
\begin{subequations}
  \label{eq:FEM}
  \begin{align}
    \label{eq:FEM-a}
    (\curl \Eth, \curl v) + (\Vt\Eth, v) +  (\grad \vph, v) 
    & = Z_h^2 (\Eth, v), && \text{ for all } v \in \Nhp,
    \\ \label{eq:FEM-b}
    (n_z^2 \vph, \psi) - (n_\tau^2 \Eth, \grad \psi)
    & = 0,  && \text{ for all } \psi \in \Vhp,
  \end{align}  
\end{subequations}
where $(\cdot, \cdot)$ denotes the (complex) inner product of
$L^2(\om)$ or its Cartesian products.

The system~\eqref{eq:FEM} leads to a generalized matrix eigenproblem
with $Z_h^2$ as the eigenvalue once a basis is used. To show it, let
$\{\psi_i: i=1, \ldots, \nV\}$ denote a basis for $\Vhp$ and let
$\{v_k: k = 1, \ldots , \nN\}$ denote a basis for $\Nhp$. Then using
the matrices
\begin{equation}
  \label{eq:17}
  \begin{aligned}
    A_{kl}
    & =     (\curl v_l, \curl v_k) + (\Vt v_l, v_k),
    &
    C_{ki}
    & = (\grad \psi_i, v_k),
    \\
    B_{il}
    & = (n_\tau^2 v_l, \grad \psi_i),
    &
    D_{ij}
    & =  -(n_z^2 \psi_j, \psi_i),
    &
    M_{kl} & = (v_l, v_k),
  \end{aligned}
\end{equation}
the discrete formulation~\eqref{eq:FEM} is equivalent to
\begin{equation}
  \label{eq:DiscreteEWP}
  \begin{bmatrix}
    A & C \\
    B & D
  \end{bmatrix}
  \begin{bmatrix}
    a \\ b 
  \end{bmatrix}
  =
  Z_h^2
  \begin{bmatrix}
    M & 0 \\
    0 & 0 
  \end{bmatrix}
  \begin{bmatrix}
    a \\ b
  \end{bmatrix}
\end{equation}
where $a$ and $b$ are vectors of (complex) coefficients in the basis
expansions
$\Eth(x, y) = \sum_k a_k v_k(x, y)$
and $ \vph(x, y) = \sum_i b_i \psi_i(x, y).$

For computation of leaky modes, the procedure is similar. The only
modification needed is that standard artificial material coefficients
dictated by a cylindrical PML \cite{monkFiniteElementMethods2003} (set e.g., in the
outermost ring of Figure~\ref{fig:kolyadin-geom}), are
incorporated in~\eqref{eq:7} (and consequently in~\eqref{eq:FEM}
and~\eqref{eq:DiscreteEWP}).
For assembling these matrices, we use the open source finite element
library NGSolve~\cite{jschoberlNGSolve} which has facilities to robustly
compute them for high values of~$p$. Our implementation of
the method is built as an add-on to this software.

\subsection{Solving the eigenproblem by FEAST algorithm}\label{subsec:FEAST}

There are many algorithmic options for solving the
constrained eigenproblem \eqref{eq:DiscreteEWP}.
In this subsection, we  show how 
an eigensolver named ``FEAST''~\cite{gopalakrishnanSpectralDiscretizationErrors2020,GuttePolizTang15,
  kestynFeastEigensolverNonHermitian2016} can be adapted for solving such 
eigenproblems.
Letting
\begin{equation}
  \label{eq:12}
  \AA = 
  \begin{bmatrix}
    A & C \\
    B & D
  \end{bmatrix}, \qquad
  \BB = 
  \begin{bmatrix}
    M & 0 \\
    0 & 0 
  \end{bmatrix},
  \qquad
  X =
  \begin{bmatrix}
    a \\ b 
  \end{bmatrix},  
\end{equation}
the eigenproblem~\eqref{eq:DiscreteEWP} for eigenvalues $\lambda=Z_h^2$
reads as the $n \times n$ generalized eigenproblem,
$\AA X = \lambda \BB X$, where
$
  n  = \nN + \nV
  = \dim(\Nhp) + \dim (\Vhp).
$
The FEAST algorithm is a technique to compute a cluster of eigenvalues, 
collected into a set $\vL \subset \C$, its
accompanying right algebraic eigenspace, denoted by $Y \subset \C^n$,
as well as the left algebraic eigenspace $\tilde Y \subset \C^n$.  The
input to the algorithm includes a simple closed contour $\vG$
which encloses the wanted cluster of eigenvalues in $\Lambda$
(without crossing any eigenvalue).
The algorithm uses approximations of Riesz projections
\begin{align}
  \label{eq:SpectralProjection}
  S=\frac{1}{2\pi\ii}\oint_\vG (z \BB - \AA)^{-1} \BB \, dz, \qquad
  \St =\frac{1}{2\pi\ii}\oint_{\vG} (z \BB - \AA)^{-*} \BB^* \, dz,
\end{align}
where $^*$ denotes the conjugate transpose.  These operators are not
generally computable, but the algorithm performs a subspace iteration
replacing them by their respective computable $N$-point quadrature
approximations,
\begin{equation}
  \label{eq:SN}
  S_N = \sum_{k=0}^{N-1} w_k (z_k  \BB  - \AA )^{-1}   \BB , 
  \qquad \St_N = \sum_{k=0}^{N-1} \bar{w}_k (z_k  \BB  - \AA )^{-*}   \BB ^*,  
\end{equation}
for carefully chosen weights $w_k \in \C$ and points $z_k \in \Gamma$.
The numbers $w_k$ and $z_k$ for circular and elliptical contours can
be found in~\cite{gopalakrishnanSpectralDiscretizationErrors2020,
 GuttePolizTang15}. Given initial
right and left subspaces $Y_0, \tilde{Y}_0 \subset \C^n$ of dimension
$m\ll n$, the algorithm computes two sequences of subspaces, $Y_\ell$ and
$\tilde{Y}_\ell$, by
\begin{equation}
  \label{eq:feast}
  Y_\ell = S_N Y_{\ell-1},
  \qquad
  \tilde{Y}_\ell = \St_N \tilde{Y}_{\ell-1},
  \qquad \text{ for } \ell = 1, 2, \ldots.  
\end{equation}
We omit a discussion of further algorithmic details, in view of the
details presented in~\cite{kestynFeastEigensolverNonHermitian2016}, except for one deviation
from the standard algorithm explained in the remainder of this
section.

In our application, the right hand side matrix $\BB$ is not
invertible. The nullspace of $\BB$ generates an eigenspace
corresponding to an infinite eigenvalue. If the subspaces $Y_\ell$ and
$\tilde{Y}_\ell$ produced by iteration~\eqref{eq:feast} contain
components in the nullspace of $\BB$, computation of eigenvalue
approximations from $Y_\ell, \tilde{Y}_\ell$ can numerically
fail. This problem can be avoided in two ways: (i) The first is to
periodically filter out the nullspace of $\BB$ while conducting the
algorithm, as described in~\cite[16-22
of Algorithm~1]{gopalakrishnanComputingLeakyModes2022}. 
(ii) A second alternative
is to compute eigenvalue approximations from $Y_\ell, \tilde{Y}_\ell$
in a nonstandard way, but leveraging the constraint equation, as
described next.

Writing the constrained eigenvalue equation~\eqref{eq:DiscreteEWP} as
a system of two equations,
\begin{align*}
  A a + C b &  = \lambda M a,  & 
  B a + D b &  = 0, 
\end{align*}
we  eliminate $b$. Since the second equation yields
$b = -D^{-1} B a$, the first equation implies
\begin{equation}
  \label{eq:11}
  T a = \lambda M a, \quad
  \text{ where }
  T = A - CD^{-1} B \in \C^{\nN \times \nN}.
\end{equation}
Suppose that at the $\ell$th
iteration in~\eqref{eq:feast}, we have a basis for the eigenspace
approximations $Y_\ell$ and $\tilde{Y}_\ell$, listed as columns of
$n \times m$ matrices $X$ and $\tilde{X}$, respectively. Block
partitioning them as in~\eqref{eq:12}, extract the top $\nN \times m$
blocks and denote them by $x, \tilde{x} \in \C^{\nN \times m}$. As the
algorithm converges, each column of $x$ and $\tilde{x}$ approximates
the $a$-block of a generalized left or right eigenvector
in~\eqref{eq:11}. It is possible to compute Ritz values using $x$ and
$\tilde x$, ignoring the remaining blocks. Since the exact eigenvalue
solves~\eqref{eq:11}, an approximate Ritz value can be obtained using
the small dense $m \times m$ matrices
\[
  T_x = \tilde x^* T
  x, \qquad
  M_x = \tilde x^* M  x.
\]
By means of standard dense linear algebra routines, we then compute
the Ritz values
$\vL_\ell = \mathrm{diag}(\lambda_1, \ldots, \lambda_m)$,
$w, \tilde w \in \C^{m \times m}$ satisfying
$\tilde w^* T_x w = \vL_\ell$ and $\tilde w^* M_x w = I$.  Unlike
$\BB$, the matrix $M$ is invertible. Hence whenever $x$ has linearly
independent columns (readily ensured in practice), $M_x$ is symmetric
and positive definite, thus making the small dense eigenproblem
robustly solvable.

\section*{Acknowledgements}

P.~Vandenberge gratefully acknowledges support from an NSF RTG Graduate
Fellowship under grant DMS-2136228. This work was also supported in
part by NSF grant DMS-1912779. 

\section*{Disclaimers}

This article is approved for public release; distribution unlimited. 
Public Affairs release approval  $\#\text{AFRL-}2023\text{-}1168$. 
The views expressed are those of the authors and do not necessarily reflect the official policy or position of the Department of the Air Force, the Department of Defense, or the U.S. government. 

\bibliography{main.bib}

\begin{thebibliography}{10}

\bibitem{belardiHollowAntiresonantFibers2014}
{\sc W.~Belardi and J.~C. Knight}, {\em Hollow antiresonant fibers with reduced
  attenuation}, Optics Letters, 39 (2014), pp.~1853--1856.

\bibitem{birdAttenuationModelHollowcore2017}
{\sc D.~Bird}, {\em Attenuation of model hollow-core, anti-resonant fibres},
  Optics Express, 25 (2017), pp.~23215--23237.

\bibitem{cocchiniLateralRigidityDoublecoated1995}
{\sc F.~Cocchini}, {\em The lateral rigidity of double-coated optical fibers},
  Journal of Lightwave Technology, 13 (1995), pp.~1706--1710.

\bibitem{duguayAntiresonantReflectingOptical1986a}
{\sc M.~A. Duguay, Y.~Kokubun, T.~L. Koch, and L.~Pfeiffer}, {\em Antiresonant
  reflecting optical waveguides in {{SiO2}}-{{Si}} multilayer structures},
  Applied Physics Letters, 49 (1986), pp.~13--15.

\bibitem{glogeOpticalfiberPackagingIts1975}
{\sc D.~Gloge}, {\em Optical-fiber packaging and its influence on fiber
  straightness and loss}, The Bell System Technical Journal, 54 (1975),
  pp.~245--262.

\bibitem{gopalakrishnanSpectralDiscretizationErrors2020}
{\sc J.~Gopalakrishnan, L.~Grubi{\v s}i{\'c}, and J.~Ovall}, {\em Spectral
  discretization errors in filtered subspace iteration}, Mathematics of
  Computation, 89 (2020), pp.~203--228.

\bibitem{gopalakrishnanComputingLeakyModes2022}
{\sc J.~Gopalakrishnan, B.~Q. Parker, and P.~VandenBerge}, {\em Computing leaky
  modes of optical fibers using a {{FEAST}} algorithm for polynomial
  eigenproblems}, Wave Motion, 108 (2022), p.~102826.

\bibitem{GuttePolizTang15}
{\sc S.~G{\"u}ttel, E.~Polizzi, P.~T.~P. Tang, and G.~Viaud}, {\em Zolotarev
  quadrature rules and load balancing for the {{FEAST}} eigensolver}, Siam
  Journal On Scientific Computing, 37 (2015), pp.~A2100--A2122.

\bibitem{hayesAntiresonantHollowCore2017}
{\sc J.~R. Hayes, S.~R. Sandoghchi, T.~D. Bradley, Z.~Liu, R.~Slav{\'i}k, M.~A.
  Gouveia, N.~V. Wheeler, G.~Jasion, Y.~Chen, E.~N. Fokoua, M.~N. Petrovich,
  D.~J. Richardson, and F.~Poletti}, {\em Antiresonant {{Hollow Core Fiber
  With}} an {{Octave Spanning Bandwidth}} for {{Short Haul Data
  Communications}}}, Journal of Lightwave Technology, 35 (2017), pp.~437--442.

\bibitem{jschoberlNGSolve}
{\sc {J Schoberl} and {et al.}}, {\em {{NGSolve}}}.

\bibitem{kestynFeastEigensolverNonHermitian2016}
{\sc J.~Kestyn, E.~Polizzi, and P.~T. Peter~Tang}, {\em Feast {{Eigensolver}}
  for {{Non-Hermitian Problems}}}, SIAM Journal on Scientific Computing, 38
  (2016), pp.~S772--S799.

\bibitem{kolyadinLightTransmissionNegative2013}
{\sc A.~N. Kolyadin, A.~F. Kosolapov, A.~D. Pryamikov, A.~S. Biriukov, V.~G.
  Plotnichenko, and E.~M. Dianov}, {\em Light transmission in negative
  curvature hollow core fiber in extremely high material loss region}, Optics
  Express, 21 (2013), pp.~9514--9519.

\bibitem{koshibaSimpleEfficientFiniteelement1992}
{\sc M.~Koshiba and K.~Inoue}, {\em Simple and efficient finite-element
  analysis of microwave and optical waveguides}, IEEE Transactions on Microwave
  Theory and Techniques, 40 (1992), pp.~371--377.

\bibitem{koshibaCurvilinearHybridEdge2000}
{\sc M.~Koshiba and Y.~Tsuji}, {\em Curvilinear hybrid edge/nodal elements with
  triangular shape for guided-wave problems}, Journal of Lightwave Technology,
  18 (2000), pp.~737--743.

\bibitem{leeFullwaveAnalysisDielectric1991}
{\sc J.-F. Lee, D.-K. Sun, and Z.~Cendes}, {\em Full-wave analysis of
  dielectric waveguides using tangential vector finite elements}, IEEE
  Transactions on Microwave Theory and Techniques, 39 (1991), pp.~1262--1271.

\bibitem{litchinitserAntiresonantReflectingPhotonic2002}
{\sc N.~M. Litchinitser, A.~K. Abeeluck, C.~Headley, and B.~J. Eggleton}, {\em
  Antiresonant reflecting photonic crystal optical waveguides}, Optics Letters,
  27 (2002), p.~1592.

\bibitem{marcuseLightTransmissionOptics1982}
{\sc D.~Marcuse}, {\em Light Transmission Optics /2nd Edition/}, {Van Nostrand
  Reinhold Company Inc.}, Jan. 1982.

\bibitem{marcuseTheoryDielectricOptical2013}
{\sc D.~Marcuse}, {\em Theory of {{Dielectric Optical Waveguides}}},
  {Elsevier}, Sept. 2013.

\bibitem{monkFiniteElementMethods2003}
{\sc P.~Monk and D.~o. M. S. P.~M. PH}, {\em Finite {{Element Methods}} for
  {{Maxwell}}'s {{Equations}}}, {Clarendon Press}, Apr. 2003.

\bibitem{monroModelingLargeAir2000}
{\sc T.~Monro, D.~Richardson, N.~Broderick, and P.~Bennett}, {\em Modeling
  large air fraction holey optical fibers}, Journal of Lightwave Technology, 18
  (2000), pp.~50--56.

\bibitem{polettiNestedAntiresonantNodeless2014}
{\sc F.~Poletti}, {\em Nested antiresonant nodeless hollow core fiber}, Optics
  Express, 22 (2014), pp.~23807--23828.

\bibitem{polettiOptimisingPerformancesHollow2011}
{\sc F.~Poletti, J.~R. Hayes, and D.~J. Richardson}, {\em Optimising the
  performances of hollow antiresonant fibres}, in 2011 37th {{European
  Conference}} and {{Exhibition}} on {{Optical Communication}}, Sept. 2011,
  pp.~1--3.

\bibitem{polyanskiyRefractiveIndexInfo}
{\sc M.~N. Polyanskiy}, {\em {{RefractiveIndex}}.{{Info}}}.
\newblock https://refractiveindex.info.

\bibitem{shiueDesignDoublecoatedOptical1992}
{\sc S.-T. Shiue}, {\em Design of double-coated optical fibers to minimize
  hydrostatic pressure induced microbending losses}, IEEE Photonics Technology
  Letters, 4 (1992), pp.~746--748.

\bibitem{snyderOpticalWaveguideTheory2024}
{\sc A.~W. Snyder and J.~Love}, {\em Optical {{Waveguide Theory}}}, {Springer
  US}, Dec. 2024.

\bibitem{snyderModesOpticalWaveguides1978}
{\sc A.~W. Snyder and W.~R. Young}, {\em Modes of optical waveguides}, JOSA, 68
  (1978), pp.~297--309.

\bibitem{uranusModellingMicrostructuredWaveguides2004}
{\sc H.~P. Uranus and H.~J. W.~M. Hoekstra}, {\em Modelling of microstructured
  waveguides using a finite-element-based vectorial mode solver with
  transparent boundary conditions}, Optics Express, 12 (2004), p.~2795.

\bibitem{vardapetyanFullWaveAnalysisDielectric2003}
{\sc L.~Vardapetyan and L.~Demkowicz}, {\em Full-{{Wave Analysis}} of
  {{Dielectric Waveguides}} at a {{Given Frequency}}}, Math. Comput., 72
  (2003), pp.~105--129.

\bibitem{yehBraggReflectionWaveguides1976}
{\sc P.~Yeh and A.~Yariv}, {\em Bragg reflection waveguides}, Optics
  Communications, 19 (1976), pp.~427--430.

\bibitem{yehTheoryBraggFiber1978}
{\sc P.~Yeh, A.~Yariv, and E.~Marom}, {\em Theory of {{Bragg Fiber}}}, JOSA, 68
  (1978), pp.~1196--1201.

\bibitem{zhangComplexRefractiveIndices2020}
{\sc X.~Zhang, J.~Qiu, X.~Li, J.~Zhao, and L.~Liu}, {\em Complex refractive
  indices measurements of polymers in visible and near-infrared bands}, Applied
  Optics, 59 (2020), pp.~2337--2344.

\end{thebibliography}
\bibliographystyle{siam}

\end{document}